\documentclass[12pt]{mychapter}

\usepackage[top=1in,left=1in,right=1in,bottom=1in]{geometry}

\newcommand{\vdd}{V$_{DD}$\xspace}
\newcommand{\halfvdd}{$\frac{1}{2}$V$_{DD}$\xspace}
\newcommand{\halfvddpd}{$\frac{1}{2}$V$_{DD}+\delta$\xspace}

\newcommand{\cmdact}{\texttt{ACTIVATE}\xspace}
\newcommand{\cmdpre}{\texttt{PRECHARGE}\xspace}
\newcommand{\cmdwr}{\texttt{WRITE}\xspace}
\newcommand{\cmdrd}{\texttt{READ}\xspace}
\newcommand{\cmdtr}{\texttt{TRANSFER}\xspace}
\newcommand{\cmdacts}{\texttt{ACTIVATE}s\xspace}

\newcommand{\bbar}{$\overline{\textrm{bitline}}$\xspace}

\newcommand{\band}{\textrm{\texttt{and}}\xspace}
\newcommand{\bor}{\textrm{\texttt{or}}\xspace}
\newcommand{\bnot}{\textrm{\texttt{not}}\xspace}
\newcommand{\bxor}{\textrm{\texttt{xor}}\xspace}
\newcommand{\bnand}{\textrm{\texttt{nand}}\xspace}
\newcommand{\bnor}{\textrm{\texttt{nor}}\xspace}
\newcommand{\bxnor}{\textrm{\texttt{xnor}}\xspace}

\newcommand{\dwordline}{\emph{d-wordline}\xspace}
\newcommand{\nwordline}{\emph{n-wordline}\xspace}

\newcommand{\baddr}[1]{\texttt{B#1}\xspace}
\newcommand{\taddr}[1]{\texttt{T#1}\xspace}
\newcommand{\daddr}[1]{\texttt{D#1}\xspace}

\newcommand{\dccdz}{\texttt{DCC0}\xspace}
\newcommand{\dccnz}{$\overline{\textrm{\texttt{DCC0}}}$\xspace}

\newcommand{\dccdo}{\texttt{DCC1}\xspace}
\newcommand{\dccno}{$\overline{\textrm{\texttt{DCC1}}}$\xspace}

\newcommand{\czero}{\texttt{C0}\xspace}
\newcommand{\cone}{\texttt{C1}\xspace}

\newcommand{\aap}{\texttt{AAP}\xspace}

\newcommand{\ap}{\texttt{AP}\xspace}

\newcommand{\tras}{t$_{\textrm{\footnotesize{RAS}}}$\xspace}
\newcommand{\trp}{t$_{\textrm{\footnotesize{RP}}}$\xspace}

\newcommand{\tra}{TRA\xspace}

\newcommand{\ambit}{Ambit\xspace}
\newcommand{\ambitao}{Ambit-AND-OR\xspace}
\newcommand{\ambitn}{Ambit-NOT\xspace}

\newcommand{\mrtwo}[1]{\multirow{2}{*}{#1}}
\newcommand{\mrthr}[1]{\multirow{3}{*}{#1}}

\usepackage{tikz}
\usetikzlibrary{arrows}
\usetikzlibrary{patterns}
\usetikzlibrary{calc}
\usetikzlibrary{shapes}
\usetikzlibrary{positioning,fit}

\usepackage{circuitikz}

\tikzset{no padding/.style={inner sep=0pt, outer sep=0pt}}
\tikzset{rounded/.style={rounded corners=2pt}}
\tikzset{rounded1/.style={rounded corners=1pt}}

\tikzset{page/.style={draw,minimum width=1.75cm,minimum height=2.25cm, rounded}}

\tikzset{value/.style={draw,minimum width=2mm,minimum height=2mm, rounded1, fill=white}}

\title{\bfseries\sffamily In-DRAM Bulk Bitwise Execution Engine}
\date{}

\author{
\begin{tabular}{ccc}
  Vivek Seshadri & & Onur Mutlu\\
  Microsoft Research India & & ETH Z{\"u}rich\\
  \texttt{visesha@microsoft.com} & & \texttt{onur.mutlu@inf.ethz.ch}
\end{tabular}
}

\begin{document}

\thispagestyle{empty}
\maketitle

\section*{Abstract}

Many applications heavily use bitwise operations on large
bitvectors as part of their computation. In existing systems,
performing such \emph{bulk bitwise operations} requires the
processor to transfer a large amount of data on the memory
channel, thereby consuming high latency, memory bandwidth, and
energy. In this paper, we describe Ambit, a recently-proposed
mechanism to perform bulk bitwise operations \emph{completely}
inside main memory. Ambit exploits the internal organization and
analog operation of DRAM-based memory to achieve low cost, high
performance, and low energy. Ambit exposes a new bulk bitwise
execution model to the host processor. Evaluations show that Ambit
significantly improves the performance of several applications
that use bulk bitwise operations, including databases.

\noindent\textbf{Index Terms:} Processing using Memory, DRAM, Bulk
Copy, Bulk Initialization, Bulk Bitwise Operations, Performance,
Energy Efficiency

\section{Introduction}

Many applications trigger \emph{bulk bitwise operations}, i.e.,
bitwise operations on large bit
vectors~\cite{btt-knuth,hacker-delight}. In databases, bitmap
indices~\cite{bmide,bmidc}, which heavily use bulk bitwise operations,
are more efficient than B-trees for many
queries~\cite{bmide,fastbit,bicompression}. In fact, many real-world
databases~\cite{oracle,redis,fastbit,rlite} support bitmap indices. A
recent work, WideTable~\cite{widetable}, designs an entire database
around a technique called BitWeaving~\cite{bitweaving}, which
accelerates \emph{scans} completely using bulk bitwise operations.
Microsoft recently open-sourced a technology called
BitFunnel~\cite{bitfunnel} that accelerates the document filtering
portion of web search. BitFunnel relies on fast bulk bitwise AND
operations. Bulk bitwise operations are also prevalent in DNA sequence
alignment~\cite{bitwise-alignment,shd,gatekeeper,grim,myers1999,nanopore-sequencing,shouji},
encryption algorithms~\cite{xor1,xor2,enc1}, graph
processing~\cite{pinatubo,pim-enabled-insts,graphpim},
networking~\cite{hacker-delight}, and machine
learning~\cite{neural-cache} . Thus, accelerating bulk bitwise
operations can significantly boost the performance of various
important applications.

In existing systems, a bulk bitwise operation requires a large amount
of data to be transferred on the memory channel. Such large data
transfers result in high latency, bandwidth, and energy
consumption. In fact, our experiments on a multi-core Intel
Skylake~\cite{intel-skylake} and an NVIDIA GeForce GTX
745~\cite{gtx745} show that the available memory bandwidth of these
systems limits the throughput of bulk bitwise operations.  Recent
works
(e.g.,~\cite{pim-enabled-insts,pim-graph,top-pim,nda,msa3d,pointer3d,tom,lazypim-cal,grim,gwcd,cds-nmc,pattnaik2016})
propose processing in the logic layer of 3D-stacked DRAM, which stacks
DRAM layers on top of a logic layer (e.g., Hybrid Memory
Cube~\cite{hmc,hmc2}, High Bandwidth Memory~\cite{hbm,smla}). While the
logic layer in 3D-stacked memory has much higher bandwidth than
traditional systems, it still cannot exploit the maximum internal
bandwidth available inside a DRAM chip~\cite{smla}
(Section~\ref{sec:lte-analysis}).

In this paper, we describe Ambit~\cite{ambit}, a new mechanism
proposed by recent work that performs bulk bitwise operation
completely inside main memory. Ambit is an instance of the
recently-introduced notion called \emph{Processing using
  Memory}~\cite{pum-bookchapter}. In contrast to Processing in Memory
architectures~\cite{pim-enabled-insts,pim-graph,top-pim,msa3d,spmm-mul-lim,data-access-opt-pim,tom,hrl,gp-simd,ndp-architecture,pim-analytics,nda,jafar,data-reorg-3d-stack,smla,lim-computer,non-von-machine,iram,execube,active-pages,pim-terasys,cram,bitwise-cal,rowclone,pointer3d,continuous-run-ahead,emc,gwcd,cds-nmc,lazypim-cal,graphpim,conda}
that add extra computational logic closer to main memory, the idea
behind Processing using Memory is to exploit the existing structure
and organization of memory devices with minimal changes to provide
additional functionality.

Along these lines, Ambit uses the analog operation principles of DRAM technology
to perform bulk bitwise operations \emph{completely inside the memory
  array}. With modest changes to the DRAM design, Ambit can exploit
1)~the maximum internal bandwidth available \emph{inside} each DRAM
array, and 2)~the memory-level parallelism~\cite{parbs,salp,glewmlp,tcm,runahead,mlp-prefetching}
\emph{across} multiple DRAM arrays to enable one to two orders of
magnitude improvement in raw throughput and energy consumption of bulk
bitwise operations. Ambit exposes a \emph{Bulk Bitwise Execution
  Model} to the host processor. We show that real-world applications
that heavily use bulk bitwise operations can use this model to achieve
significant improvements in performance and energy efficiency.

In this paper, we discuss the following main concepts.
\begin{itemize}
\item As Ambit builds on top of modern DRAM architecture, we first
  provide a brief background on modern DRAM organization and
  operation that is sufficient to understand the mechanisms
  proposed by Ambit (Section~\ref{sec:background}).
\item We describe the different components of Ambit, their design and
  implementation, and execution models to expose Ambit to the host
  system in detail (Sections~\ref{sec:ambit}--\ref{sec:support}).
\item We describe quantitative evaluations showing that 1)~Ambit works
  reliably even under significant process variation, and 2)~improves
  performance and energy-efficiency compared to existing systems
  for real workloads
  (Sections~\ref{sec:spice-sim}--\ref{sec:applications}).
\end{itemize}

\section{Background on DRAM}
\label{sec:background}

In this section, we describe the necessary background to understand
modern DRAM architecture and its implementation. This paper builds
on our previous book chapter that introduces the notion of
\emph{Processing using Memory}~\cite{pum-bookchapter}. Since that
chapter provides a detailed background on DRAM, this section is mostly
reproduced from the DRAM background section from chapter. While we
focus our attention primarily on commodity DRAM design (i.e., the DDRx
interface), most DRAM architectures use very similar design approaches
and vary only in higher-level design choices~\cite{ramulator}. As a
result, Ambit can be extended to any DRAM architecture. There has been
significant recent research in DRAM architectures and the interested
reader can find details about various aspects of DRAM in multiple
recent
publications~\cite{salp,tl-dram,al-dram,gs-dram,dsarp,ramulator,data-retention,parbor,fly-dram,efficacy-error-techniques,raidr,chargecache,avatar,diva-dram,reaper,softmc,lisa,smla,vivek-thesis,yoongu-thesis,donghyuk-thesis,kevin-thesis,rowclone,ddma,samira-micro17,samira-cal16,crow,cal-dram,drange,dlpuf,solar-dram,patel2019}.

At the end of this section, we provide a brief overview of
RowClone~\cite{rowclone}, a prior work that enables the memory
controller to perform row-wide copy and initialization operations
completely inside DRAM. Ambit exploits RowClone to reduce the overhead
of some of its bulk data copy and initialization operations.

\subsection{High-level Organization of the Memory System}

\begin{figure}[h]
  \centering
  \includegraphics{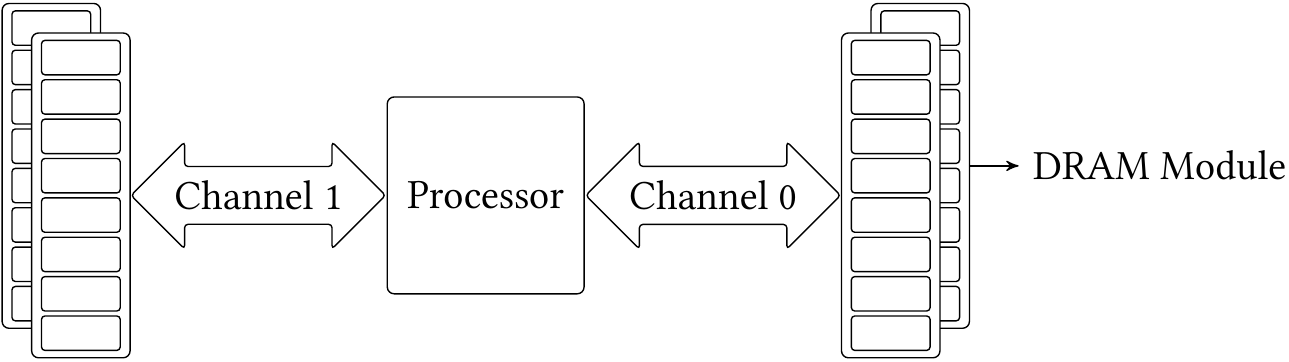}
  \caption{High-level organization of the memory subsystem}
  \label{fig:high-level-mem-org}
\end{figure}

Figure~\ref{fig:high-level-mem-org} shows the organization of the
memory subsystem in a modern computing system. At a high level, each
processor chip is connected to of one of more off-chip memory
\emph{channels}. Each memory channel consists of its own set of
\emph{command}, \emph{address}, and \emph{data} buses. Depending on
the design of the processor, there can be either an independent memory
controller for each memory channel or a single memory controller for
all memory channels. All memory modules connected to a channel share
the buses of the channel. Each memory module consists of many DRAM
devices (or chips). Most of this section is dedicated to describing
the design of a modern DRAM chip. In Section~\ref{sec:dram-module}, we
present more details of the module organization of commodity DRAM.

\subsection{DRAM Chip}
\label{sec:dram-chip}

A modern DRAM chip consists of a hierarchy of structures: DRAM
\emph{cells}, \emph{tiles/MATs}, \emph{subarrays}, and
\emph{banks}. In this section, we  describe the design of a
modern DRAM chip in a bottom-up fashion, starting from a single
DRAM cell and its operation.

\subsubsection{DRAM Cell and Sense Amplifier}

\begin{figure}[h]
  \centering
  \includegraphics{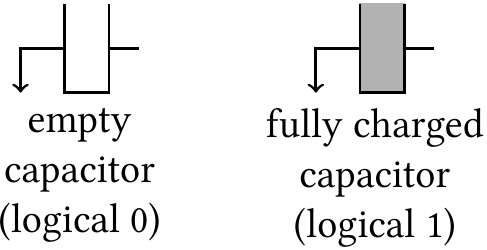}
  \caption[Capacitor]{Two states of a capacitor}
  \label{fig:cell-states}
\end{figure}

At the lowest level, DRAM technology uses capacitors to store
information. Specifically, it uses the two extreme states of a
capacitor, namely, the \emph{empty} and the \emph{fully charged}
states to store a single bit of information. For instance, an
empty capacitor can denote a logical value of 0, and a fully
charged capacitor can denote a logical value of 1.
Figure~\ref{fig:cell-states} shows the two extreme states of a
capacitor.

Unfortunately, the capacitors used for DRAM chips are small, and
will get smaller with each new generation. As a result, the amount
of charge that can be stored in the capacitor, and hence the
difference between the two states is also very small. In addition,
the capacitor can potentially lose its state after it is
accessed. Therefore, to extract the state of the capacitor, DRAM
manufacturers use a component called \emph{sense amplifier}.

\begin{figure}[h]
  \centering
  \begin{minipage}{5cm}
    \centering
    \includegraphics{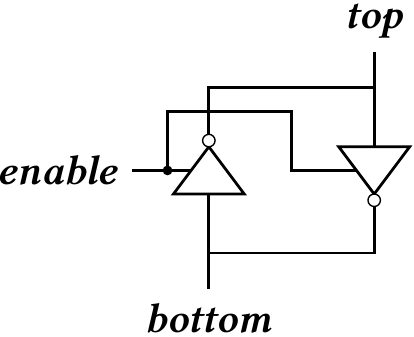}
    \caption{Sense amplifier}
    \label{fig:sense-amp}
  \end{minipage}\quad
  \begin{minipage}{9cm}
    \centering
    \includegraphics{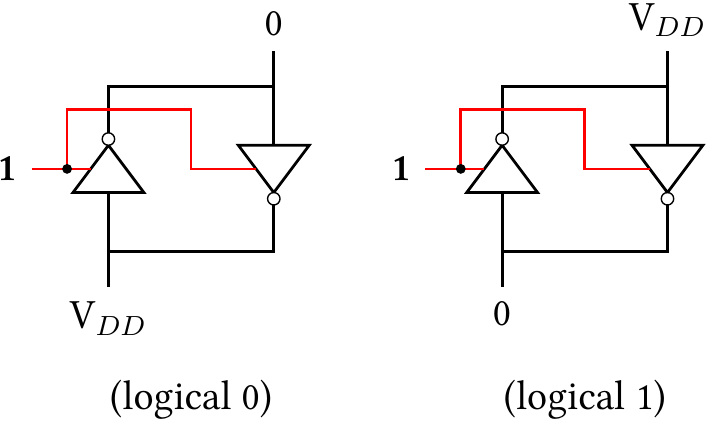}
    \caption{Stable states of a sense amplifier}
    \label{fig:sense-amp-states}
  \end{minipage}
\end{figure}

Figure~\ref{fig:sense-amp} shows a sense amplifier. A sense
amplifier contains two inverters which are connected together such
that the output of one inverter is connected to the input of the
other and vice versa. The sense amplifier also has an enable
signal that determines if the inverters are active. When enabled,
the sense amplifier has two stable states, as shown in
Figure~\ref{fig:sense-amp-states}. In both these stable states,
each inverter takes a logical value and feeds the other inverter
with the negated input.

\begin{figure}[h]
  \centering
  \includegraphics{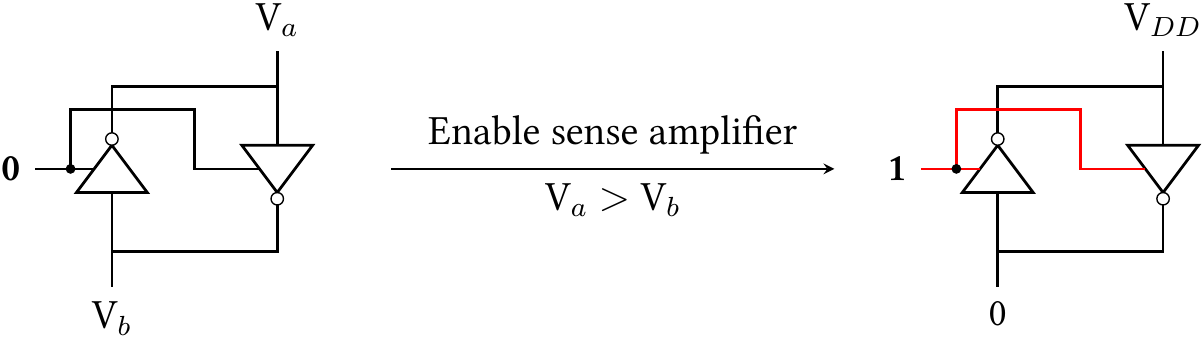}
  \caption{Operation of the sense amplifier}
  \label{fig:sense-amp-operation}
\end{figure}

Figure~\ref{fig:sense-amp-operation} shows the operation of the sense
amplifier, starting from a disabled state. In the initial disabled
state, we assume that the voltage level of the top terminal (V$_a$) is
higher than that of the bottom terminal (V$_b$).  When the sense
amplifier is enabled in this state, it \emph{senses} the difference
between the two terminals and \emph{amplifies} the difference until it
reaches one of the stable states (hence the name ``sense amplifier'').

\subsubsection{DRAM Cell Operation: The \texttt{ACTIVATE-PRECHARGE}
  cycle}
\label{sec:cell-operation}

\begin{figure}[h]
  \centering
  \begin{tikzpicture}
    \node (cop) {
      \includegraphics{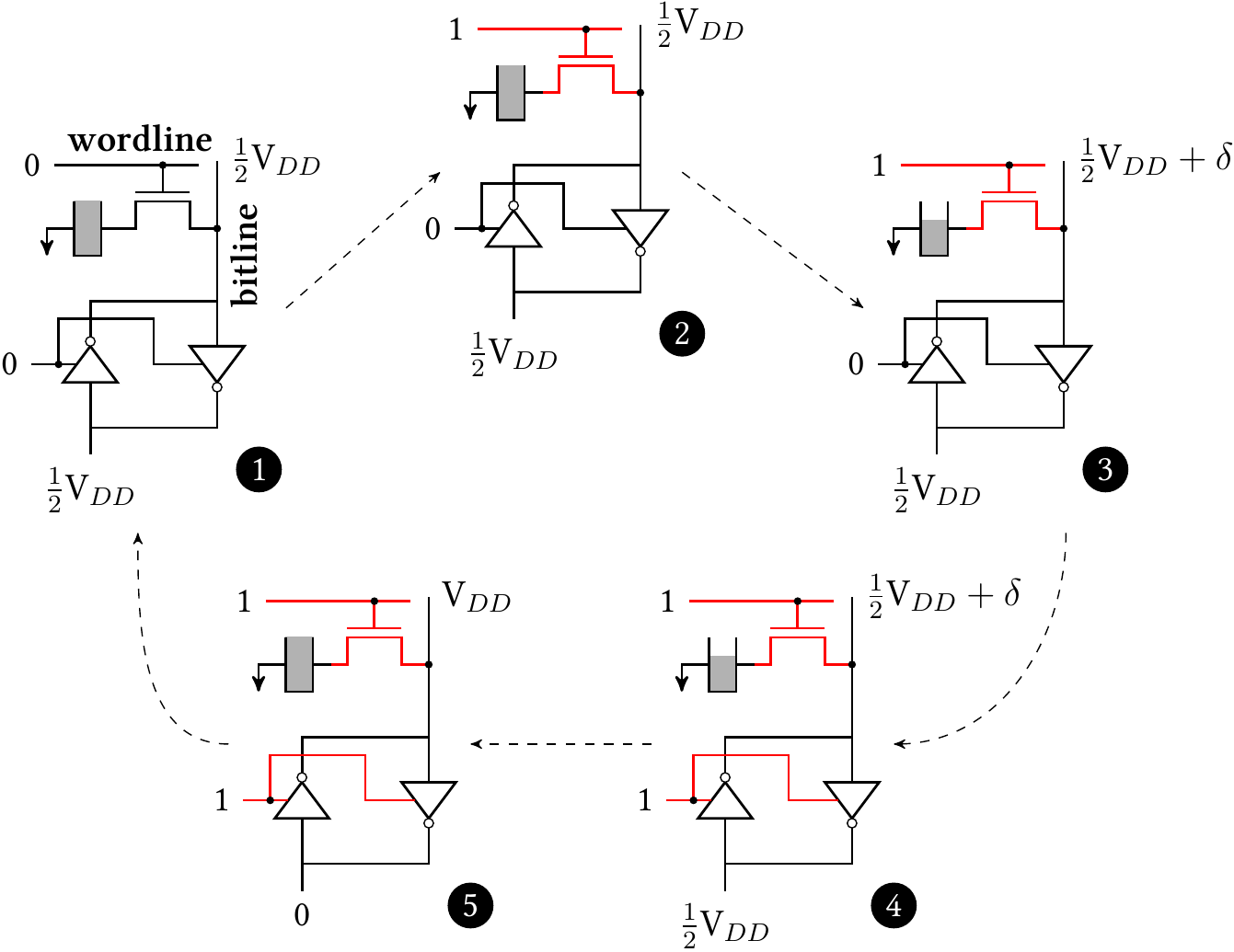}
    };
    \node at (cop) [xshift=-28mm,yshift=30mm,rotate=40] {\cmdact};
    \node at (cop) [xshift=-65mm,yshift=-18mm,rotate=0] {\cmdpre};
  \end{tikzpicture}
  \caption{Operation of a DRAM cell and sense amplifier. \ding{202}
    precharged state, \ding{203} wordline enable, \ding{204} charge
    sharing, \ding{205} sense amplifier enable, \ding{206} activated
    state.}
  \label{fig:cell-operation}
\end{figure}

DRAM technology uses a simple mechanism that converts the logical
state of a capacitor into a logical state of the sense amplifier. Data
can then be accessed from the sense amplifier (since it is in a stable
state). Figure~\ref{fig:cell-operation} shows 1)~the connection
between a DRAM cell and the sense amplifier, and 2)~the sequence of
states involved in capturing the cell state in the sense amplifier.

As shown in the figure (state \ding{202}), the capacitor is connected
to an access transistor that acts as a switch between the capacitor
and the sense amplifier. The transistor is controller by a wire called
\emph{wordline}. The wire that connects the transistor to the top end
of the sense amplifier is called \emph{bitline}. In the initial state
\ding{202}, the wordline is lowered, the sense amplifier is disabled
and both ends of the sense amplifier are maintained at a voltage level
of \halfvdd. We assume that the capacitor is initially fully charged
(the operation is similar if the capacitor was initially empty). This
state is referred to as the \emph{precharged} state. An access to the
cell is triggered by a command called \cmdact. Upon receiving an
\cmdact, the corresponding wordline is first raised (state
\ding{203}). This connects the capacitor to the bitline. In the
ensuing phase called \emph{charge sharing} (state \ding{204}), charge
flows from the capacitor to the bitline, raising the voltage level on
the bitline (top end of the sense amplifier) to \halfvddpd. After
charge sharing, the sense amplifier is enabled (state \ding{205}). The
sense amplifier detects the difference in voltage levels between its
two ends and amplifies the deviation, till it reaches the stable state
where the top end is at \vdd (state \ding{206}). Since the capacitor
is still connected to the bitline, the charge on the capacitor is also
fully restored. We shortly describe how the data can be accessed from
the sense amplifier. However, once the access to the cell is complete,
the cell is taken back to the original precharged state using the
command called \cmdpre. Upon receiving a \cmdpre, the wordline is
first lowered, thereby disconnecting the cell from the sense
amplifier. Then, the two ends of the sense amplifier are driven to
\halfvdd using a precharge unit (not shown in the figure for brevity).

\subsubsection{DRAM MAT/Tile: The Open Bitline Architecture}
\label{sec:dram-mat}

A major goal of DRAM manufacturers is to maximize the density of
the DRAM chips while adhering to certain latency constraints
(described in Section~\ref{sec:dram-timing-constraints}). There
are two costly components in the setup described in the previous
section. The first component is the sense amplifier itself. Each
sense amplifier is around two orders of magnitude larger than a
single DRAM cell~\cite{rambus-power,tl-dram}. Second, the state of the
wordline is a function of the address that is currently being
accessed. The logic that is necessary to implement this function
(for each cell) is expensive.

\begin{figure}[h]
  \centering
  \includegraphics{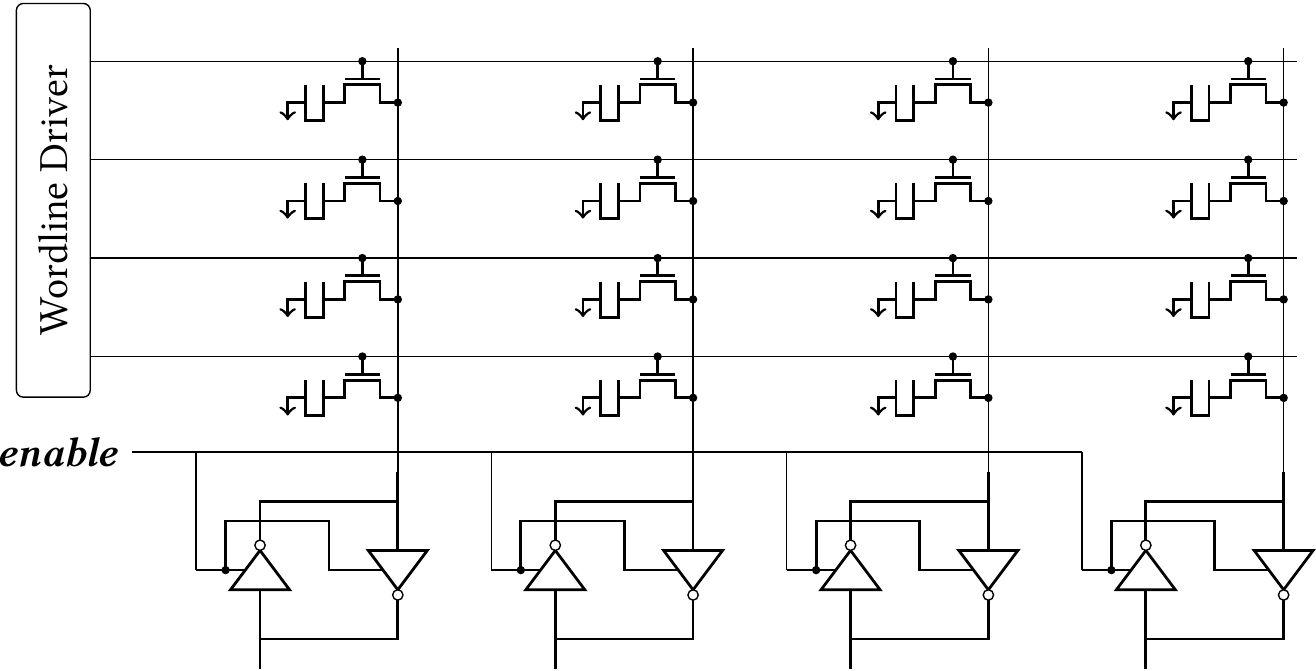}
  \caption{A 2-D array of DRAM cells}
  \label{fig:dram-mat}
\end{figure}

In order to reduce the overall cost of these two components, they
are shared by many DRAM cells. Specifically, each sense amplifier
is shared by a column of DRAM cells. In other words, all the cells in
a single column are connected to the same bitline. Similarly, each
wordline is shared by a row of DRAM cells. Together, this
organization consists of a 2-D array of DRAM cells connected to a
row of sense amplifiers and a column of wordline
drivers. Figure~\ref{fig:dram-mat} shows this organization with a
$4 \times 4$ 2-D array.

\begin{figure}[h]
  \centering
  \includegraphics{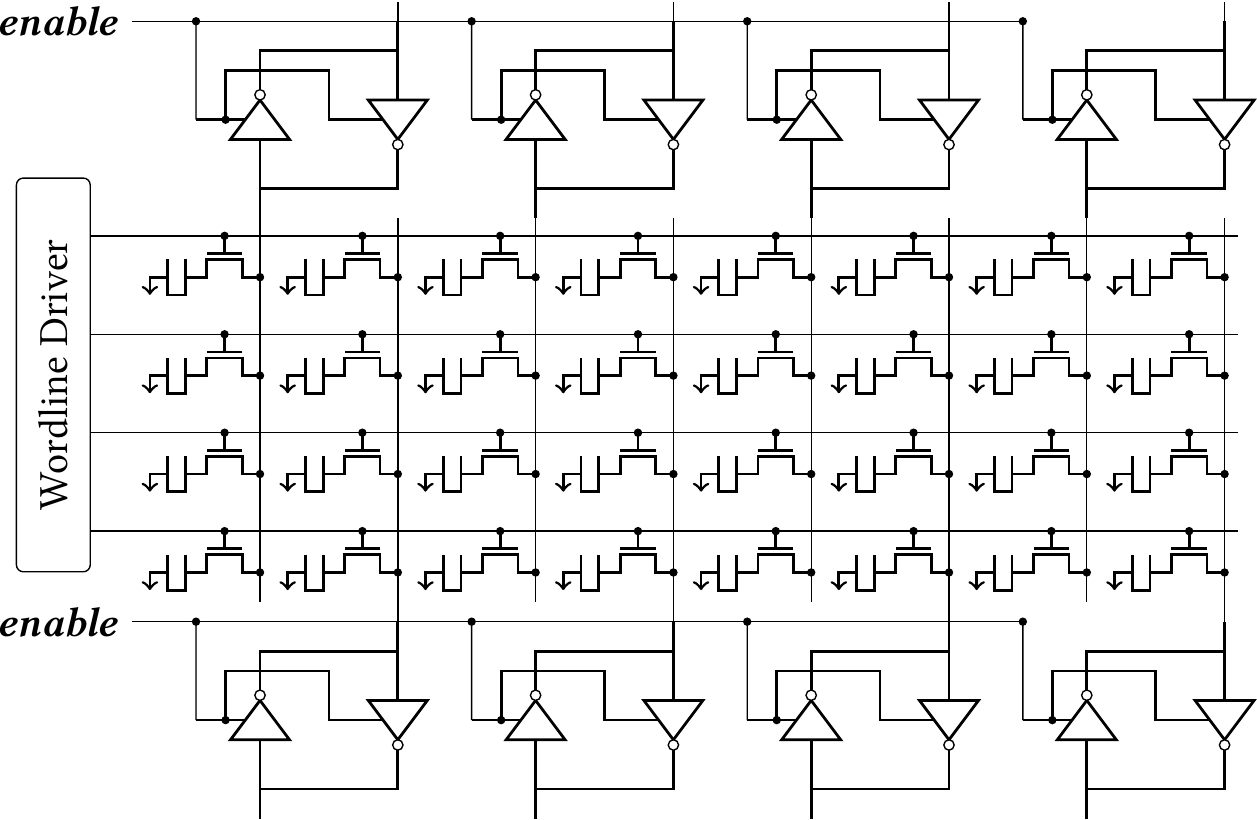}
  \caption{A DRAM MAT/Tile: The open bitline architecture}
  \label{fig:dram-mat-oba}
\end{figure}

To further reduce the overall cost of the sense amplifiers and the
wordline driver, modern DRAM chips use an architecture called the
\emph{open bitline architecture}~\cite{lisa}. This architecture
exploits two observations. First, the sense amplifier is wider than
the DRAM cells. This difference in width results in a white space near
each column of cells. Second, the sense amplifier is
symmetric. Therefore, cells can also be connected to the bottom part
of the sense amplifier. Putting together these two observations, we
can pack twice as many cells in the same area using the open bitline
architecture, as shown in Figure~\ref{fig:dram-mat-oba};

As shown in the figure, a 2-D array of DRAM cells is connected to
two rows of sense amplifiers: one on the top and one on the bottom
of the array. While all the cells in a given row share a common
wordline, half the cells in each row are connected to the top row
of sense amplifiers and the remaining half of the cells are
connected to the bottom row of sense amplifiers. This tightly
packed structure is called a DRAM
MAT/Tile~\cite{rethinking-dram,half-dram,salp}. In a modern DRAM
chip, each MAT typically is a $512 \times 512$ or $1024 \times
1024$ array. Multiple MATs are grouped together to form a larger
structure called a \emph{DRAM bank}, which we describe next.

\subsubsection{DRAM Bank}

\begin{figure}
  \centering
  \input{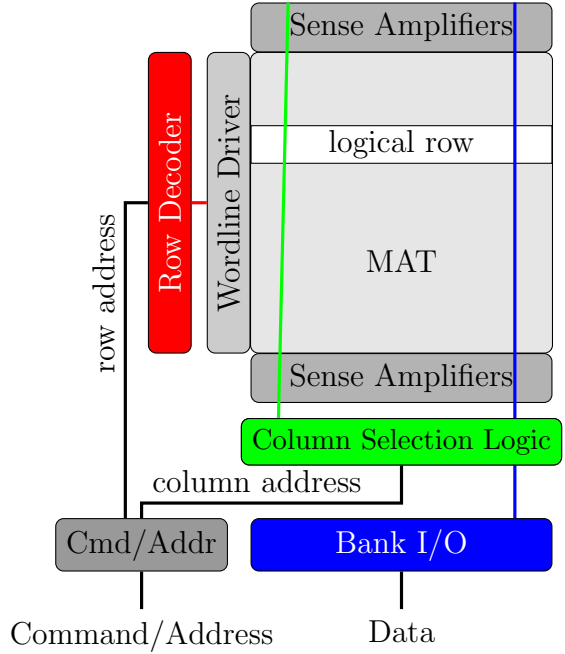}
  \caption{DRAM Bank: Logical view}
  \label{fig:dram-bank-logical}
\end{figure}

In most modern commodity DRAM interfaces~\cite{ddr3,ddr4,ramulator}, a
DRAM bank is the smallest structure visible to the memory
controller. All commands related to data access are directed to a
specific bank. Logically, each DRAM bank is a large monolithic
structure with a 2-D array of DRAM cells connected to a single set of
sense amplifiers (also referred to as a row buffer). For example, in a
2Gb DRAM chip with 8 banks, each bank has $2^{15}$ rows and each
logical row has 8192 DRAM cells. Figure~\ref{fig:dram-bank-logical}
shows this logical view of a bank.

In addition to the MAT, the array of sense amplifiers, and the
wordline driver, each bank also consists of some peripheral structures
to decode DRAM commands and addresses, and manage the inputs/outputs
to the DRAM bank. Specifically, each bank has a \emph{row decoder} to
decode the \emph{row address} associated with row-level commands
(e.g., \cmdact). Each data access command (\cmdrd and \cmdwr) accesses
only a part of a DRAM row. Such an individual part is referred to as a
\emph{column}. With each data access command, the address of the
column to be accessed is provided. This address is decoded by the
\emph{column selection logic}. Depending on which column is selected,
the corresponding piece of data is communicated between the sense
amplifiers and the \emph{bank I/O logic}. The bank I/O logic in turn
acts as an interface between the DRAM bank and the \emph{chip-level
  I/O logic}.

\begin{figure}[h]
  \centering
  \includegraphics[scale=0.9]{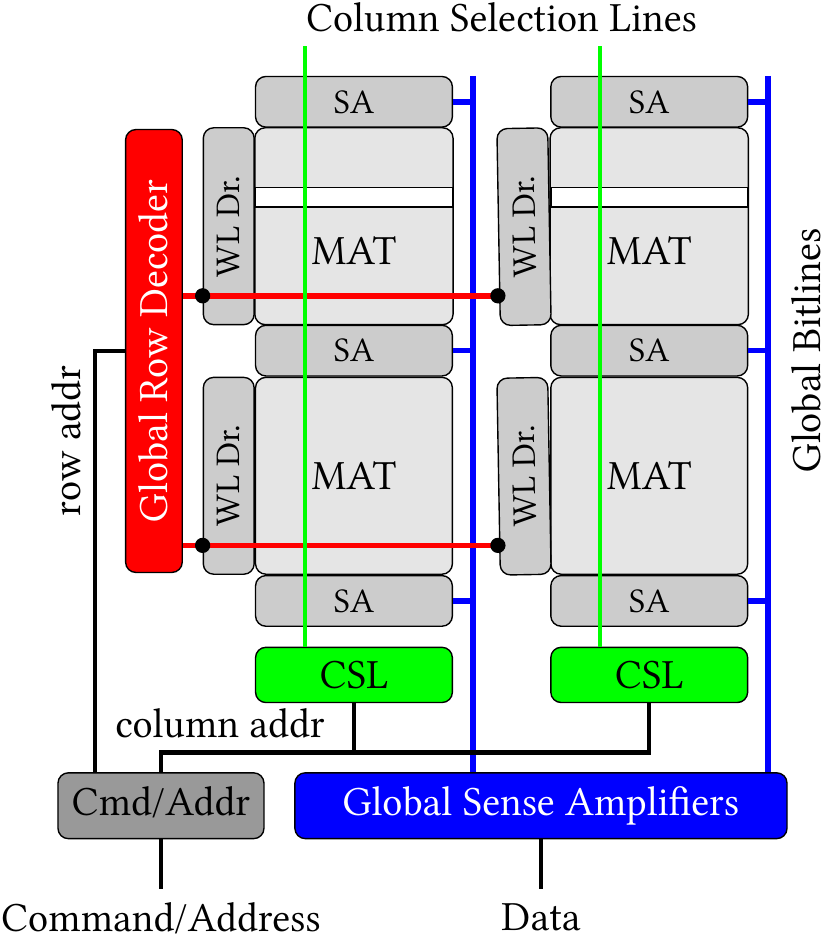}
  \caption{DRAM Bank: Physical implementation. The global bitlines
    run on top of the MATs in a separate metal layer. (components
    in the figure are not to scale)}
  \label{fig:dram-bank-physical}
\end{figure}

Although the bank can logically be viewed as a single MAT,
building a single MAT of a very large size is practically not
feasible as it would require very long bitlines and wordlines
(leading to very high latencies). Therefore, each bank is
physically implemented as a 2-D array of DRAM
MATs. Figure~\ref{fig:dram-bank-physical} shows a physical
implementation of the DRAM bank with 4 MATs arranged in a $2
\times 2$ array. As shown in the figure, the output of the global
row decoder is sent to each row of MATs.  The bank I/O logic, also
known as the \emph{global sense amplifiers}, are connected to all
the MATs through a set of \emph{global bitlines}. As shown in the
figure, each vertical collection of MATs has its own column
selection logic (CSL) and global bitlines. In a real DRAM chip,
the global bitlines run on top of the MATs in a separate metal
layer. Data of each column is split equally across a single row of
MATs. With this data organization, each global bitline needs to be
connected only to bitlines within one MAT. While prior
work~\cite{rethinking-dram} has explored routing mechanisms to
connect each global bitline to all MATs, such a design incurs high
complexity and overhead.

\begin{figure}[h!]
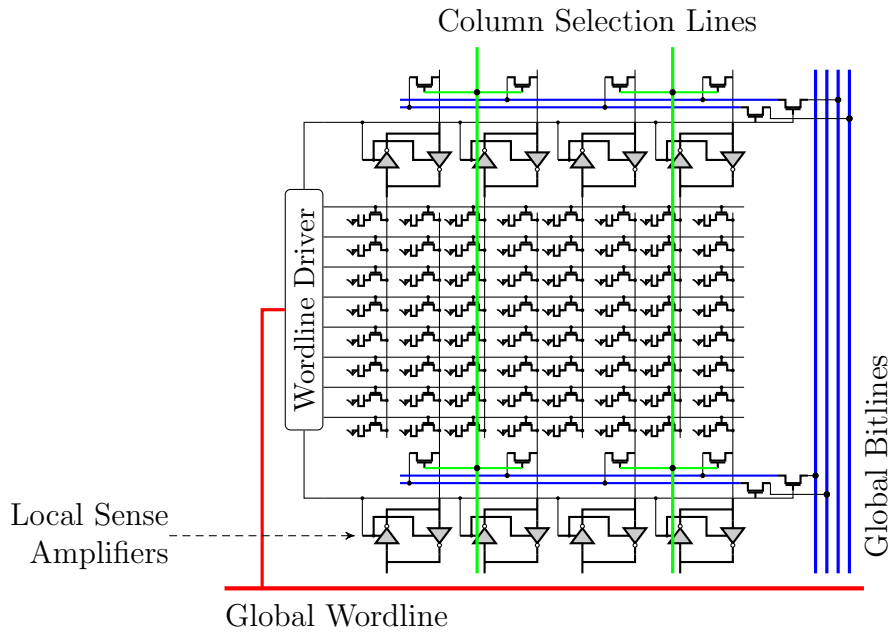

  \centering
  \begin{tikzpicture}[rounded corners=0pt,inner sep=0pt,anchor=center]

  \input{tikz-styles}
  \input{dram-stuff}

  \coordinate (origin) at (0,0);

  \foreach \i in {0,1,2,3} {
    \coordinate (sa1-loc-\i) at ([xshift=1.3*\i cm]origin);
    \coordinate (sa2-loc-\i) at ([yshift=5cm]sa1-loc-\i);

    % sense amps
    \senseampfull{sa1-\i}{sa1-loc-\i}{}{0.5};
    \senseampfull{sa2-\i}{sa2-loc-\i}{}{0.5};

    % cells
    \coordinate (cell-start-\i) at ([yshift=9mm]sa1-\i-top);
    \foreach \j in {0,...,7} {
      \coordinate (cell-contact-\i\j) at ([yshift=0.4*\j cm]cell-start-\i);
      \dramcell{cell-\i\j}{cell-contact-\i\j}{0}{}{0.25};
      \draw [fill] (cell-contact-\i\j) circle (0.5pt);
      \draw [fill] (cell-\i\j-trans-gate) circle (0.5pt);

      \coordinate (cell2-contact-\i\j) at (cell-contact-\i\j-|sa2-\i-bottom);
      \dramcell{cell2-\i\j}{cell2-contact-\i\j}{0}{}{0.25};
      \draw [fill] (cell2-contact-\i\j) circle (0.5pt);
      \draw [fill] (cell2-\i\j-trans-gate) circle (0.5pt);
    }

    % bitlines
    \draw (sa1-\i-top) -- (cell-contact-\i7) -- ++(0,1mm);
    \draw (sa2-\i-bottom) -- (cell2-contact-\i0) -- ++(0,-1mm);
    \draw (sa2-\i-top) -- ++(0,7mm);
  }

  % wordline driver
  \node (wordline-driver) at ([xshift=-1.8cm]cell-start-0) [anchor=west,rotate=90,minimum
    width=3.2cm,draw,rounded corners=2pt,minimum height=0.5cm]
        {\small Wordline Driver};

        % wordlines
        \foreach \i in {0,...,7} {
          \draw (wordline-driver.south|-cell-0\i-trans-gate) --
          (cell-0\i-trans-gate) -- (cell-3\i-trans-gate) -- ++(3mm,0);
        }

        % enable signal
        \coordinate (enable-signal1) at ([xshift=-5mm]sa1-0-enable|-sa1-0-top);
        \draw (enable-signal1) -- (enable-signal1-|sa1-3-enable);
        \foreach \i in {0,1,2,3} {
          \draw [fill] (enable-signal1-|sa1-\i-enable) circle (0.5pt) -- (sa1-\i-enable);
        }

        \coordinate (enable-signal2) at ([xshift=-5mm]sa2-0-enable|-sa2-0-top);
        \draw (enable-signal2) -- (enable-signal2-|sa2-3-enable);
        \foreach \i in {0,1,2,3} {
          \draw [fill] (enable-signal2-|sa2-\i-enable) circle (0.5pt) -- (sa2-\i-enable);
        }

        \draw (wordline-driver.east) |- (enable-signal2);
        \draw (wordline-driver.west) |- (enable-signal1);

        \foreach \j in {1,2} {
          % global bitline connection
          \coordinate (gbl-con\j1) at ([yshift=2mm,xshift=1cm]enable-signal\j);
          \coordinate (gbl-con\j1-trans-loc) at ([xshift=5mm]sa\j-3-top|-gbl-con\j1);
          \transbottom{gbl-con\j1-trans}{gbl-con\j1-trans-loc}{1};
          \draw [thick,blue] (gbl-con\j1) -- (gbl-con\j1-trans-contact);

          \coordinate (gbl-con\j2) at ([yshift=1mm]gbl-con\j1);
          \coordinate (gbl-con\j2-trans-loc) at ([xshift=5mm]gbl-con\j1-trans-loc|-gbl-con\j2);
          \transbottom{gbl-con\j2-trans}{gbl-con\j2-trans-loc}{1};
          \draw [thick,blue] (gbl-con\j2) -- (gbl-con\j2-trans-contact);

          % gbl connection trans enable
          \draw (enable-signal\j-|sa\j-3-enable) -| (gbl-con\j1-trans-gate);
          \draw (enable-signal\j-|gbl-con\j1-trans-gate) -| (gbl-con\j2-trans-gate);
          
          \foreach \i in {0,1,2,3} {
            \coordinate (data-trans-loc-\j\i) at ([yshift=6mm]sa\j-\i-top);
            \transbottom{data-trans-\j\i}{data-trans-loc-\j\i}{1};
          }

          \foreach \i in {0,2} {
            \draw [fill] (data-trans-\j\i-contact) --
            (data-trans-\j\i-contact|-gbl-con\j1) circle (0.75pt);
          }      
          \foreach \i in {1,3} {
            \draw [fill] (data-trans-\j\i-contact) --
            (data-trans-\j\i-contact|-gbl-con\j2) circle (0.75pt);
          }      
        }

        % global bit lines
        \coordinate (gbl1-start) at ([xshift=1mm]gbl-con12-trans-loc|-sa1-3-bottom);
        \coordinate (gbl1-end) at ([yshift=7mm]sa2-3-top-|gbl1-start);

        \foreach \i/\j in {1/2,2/3,3/4} {
          \coordinate (gbl\j-start) at ([xshift=1.5mm]gbl\i-start);
          \coordinate (gbl\j-end) at ([xshift=1.5mm]gbl\i-end);
        }

        \foreach \i in {1,2,3,4} {
          \draw [very thick,blue] (gbl\i-start) -- (gbl\i-end);
        }

        % gbl conn bit line connections
        \draw [fill] (gbl-con12-trans-loc) -- (gbl-con12-trans-loc-|gbl1-start) circle(1pt);
        \draw (gbl-con11-trans-loc) |- ([yshift=-1.5mm]gbl-con11-trans-loc-|gbl2-start) circle(1pt);
        \draw [fill] ([yshift=-1.5mm]gbl-con11-trans-loc-|gbl2-start) circle(1pt);
        \draw [fill] (gbl-con22-trans-loc) -- (gbl-con22-trans-loc-|gbl3-start) circle(1pt);
        \draw (gbl-con21-trans-loc) |- ([yshift=-1.5mm]gbl-con21-trans-loc-|gbl4-start) circle(1pt);
        \draw [fill] ([yshift=-1.5mm]gbl-con21-trans-loc-|gbl4-start) circle(1pt);

        % column selection
        \coordinate (colsel1) at ([xshift=5.5mm]$(sa1-0-bottom)!0.5!(sa1-1-bottom)$);
        \draw [very thick,green!100] (colsel1) -- ([yshift=10mm]colsel1|-sa2-0-top);
        
        \coordinate (colsel2) at ([xshift=5.5mm]$(sa1-2-bottom)!0.5!(sa1-3-bottom)$);
        \draw [very thick,green] (colsel2) -- ([yshift=10mm]colsel2|-sa2-0-top)
          node [anchor=south,black,yshift=2mm,xshift=-1cm] {Column Selection Lines};
        
        \foreach \i in {1,2} {
          \draw [thick,green] (data-trans-\i0-gate) -- (data-trans-\i1-gate);
          \draw [fill] (data-trans-\i0-gate-|colsel1) circle(1pt);
          \draw [thick,green] (data-trans-\i2-gate) -- (data-trans-\i3-gate);
          \draw [fill] (data-trans-\i2-gate-|colsel2) circle(1pt);
        }

        % global word line
        \coordinate (gwl-start) at ([xshift=-3mm,yshift=-2mm]wordline-driver.north|-sa1-0-bottom);
        \draw [ultra thick,red] (gwl-start) +(-0.5,0) -- +(8,0);
        \draw [very thick,red]  (gwl-start) |- (wordline-driver.north);

        \node at ([xshift=-5mm,yshift=-2mm]gwl-start) [anchor=north west] {Global Wordline};
        \node at ([xshift=2mm,yshift=2mm]gbl4-start) [anchor=north west,rotate=90] {Global Bitlines};

        \node (salabel) at (origin) [text width=2.5cm,xshift=-4.5cm,align=right] {Local Sense Amplifiers};
        \draw [->,>=stealth',densely dashed] (salabel.east) -- ([xshift=-1mm]sa1-0-enable);
\end{tikzpicture}
  \caption{Detailed view of a MAT}
  \label{fig:dram-mat-zoomed}
\end{figure}

Figure~\ref{fig:dram-mat-zoomed} shows the zoomed-in version of a
DRAM MAT with the surrounding peripheral logic. Specifically, the
figure shows how each column selection logic selects specific
sense amplifiers from a MAT and connects them to the global
bitlines. We note that the width of the global bitlines for each
MAT (typically 8/16) is much smaller than that of the width of the
MAT (typically 512/1024). This is because the global bitlines span
a much longer distance across the chip and hence have to be wider
to ensure signal integrity.

\begin{figure}[h]
  \centering
  \includegraphics{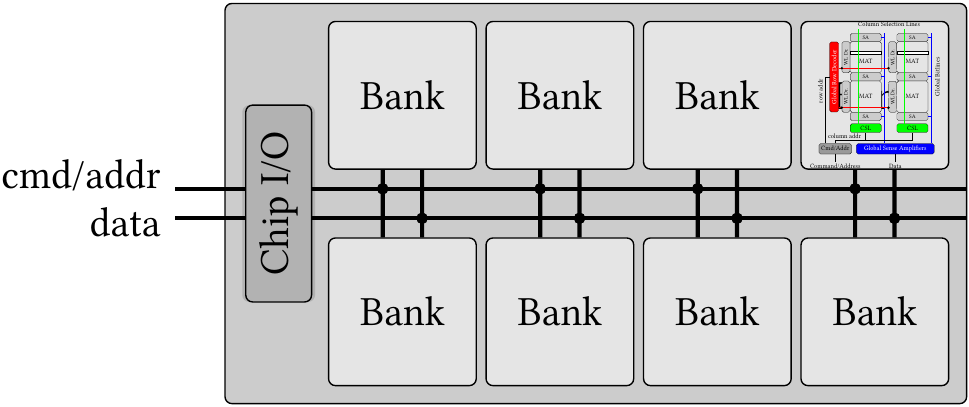}
  \caption{DRAM Chip}
  \label{fig:dram-chip}
\end{figure}

Each DRAM chip consists of multiple banks, as shown in
Figure~\ref{fig:dram-chip}. All the banks share the chip's internal
command, address, and data buses. As mentioned before, each bank
operates mostly independently (except for operations that involve the
shared buses). The chip I/O logic manages the transfer of data to and
from the chip's internal bus to the memory channel. The width of the
chip output (typically 8 bits) is much smaller than the output width
of each bank (typically 64 bits). Any piece of data accessed from a
DRAM bank is first buffered at the chip I/O and sent out on the memory
bus 8 bits at a time. With the DDR (double data rate) technology, 8
bits are sent out each half cycle. Therefore, it takes 4 cycles to
transfer 64 bits of data from a DRAM chip I/O logic on to the memory
channel.

\subsubsection{DRAM Commands: Accessing Data from a DRAM Chip}

To access a piece of data from a DRAM chip, the memory controller
must first identify the location of the data: the bank ID ($B$),
the row address ($R$) within the bank, and the column address
($C$) within the row. After identifying these pieces of
information, accessing the data involves three steps.

The first step is to issue a \cmdpre to the bank $B$. This step
prepares the bank for a data access by ensuring that all the sense
amplifiers are in the \emph{precharged} state
(Figure~\ref{fig:cell-operation}, state~\ding{202}). No wordline
within the bank is raised in this state.

The second step is to activate the row $R$ that contains the
data. This step is triggered by issuing an \cmdact to bank $B$ with
row address $R$. Upon receiving this command, the corresponding
bank feeds its global row decoder with the input $R$. The global
row decoder logic then raises the wordline of the DRAM row
corresponding to address $R$ and enables the sense amplifiers
connected to that row. This triggers the DRAM cell operation
described in Section~\ref{sec:cell-operation}. At the end of the
activate operation, the data from the entire row of DRAM cells is
copied to the corresponding array of sense amplifiers.

The third and final step is to access the data from the required
column. This is done by issuing a \cmdrd or \cmdwr command to the bank
with the column address $C$. Upon receiving a \cmdrd or \cmdwr
command, the corresponding address is fed to the column selection
logic. The column selection logic then raises the column selection
lines (Figure~\ref{fig:dram-mat-zoomed}) corresponding to address $C$,
thereby connecting those local sense amplifiers to the global sense
amplifiers through the global bitlines. For a read access, the global
sense amplifiers sense the data from the MAT's local sense amplifiers
and transfer that data to the chip's internal bus. For a write access,
the global sense amplifiers read the data from the chip's internal bus
and force the MAT's local sense amplifiers to the appropriate state.

Not all data accesses require all three steps. Specifically, if the
row that is to be accessed is already activated in the corresponding
bank, then the first two steps can be skipped and the data can be
directly accessed by issuing a \cmdrd or \cmdwr to the bank. For this
reason, the array of sense amplifiers are also referred to as a
\emph{row buffer}, and such an access that skips the first two steps
is called a \emph{row buffer hit}. Similarly, if the bank is already
in the precharged state, then the first step can be skipped. Such an
access is referred to as a \emph{row buffer miss}. Finally, if a
different row is activated within the bank, then all three steps have
to be performed. Such an access is referred to as a \emph{row buffer
  conflict}.

\subsubsection{DRAM Timing Constraints}
\label{sec:dram-timing-constraints}

Different operations within DRAM consume different amounts of
time. Therefore, after issuing a command, the memory controller must
wait for a sufficient amount of time before it can issue the next
command. Such wait times are managed by pre-specified fixed delays,
called the \emph{timing constraints}. Timing constraints essentially
dictate the minimum amount of time between two commands issued to the
same bank/rank/channel. Table~\ref{table:timing-constraints} describes
some key timing constraints along with their values for the DDR3-1600
interface. The reason as to why these constraints exist is discussed
in prior works~\cite{salp,tl-dram,chang2017,smc}.

\begin{table}[h]\small
  \centering
  \begin{tabular}{lrclp{2.2in}r}
  \toprule
  Name & \multicolumn{3}{c}{Constraint} & Description & Value (ns)\\
  \toprule
  \mrtwo{tRAS} & \mrtwo \cmdact & \mrtwo{$\rightarrow$} & \mrtwo \cmdpre & Time taken to complete a row
  activation operation in a bank & \mrtwo{35}\\
  \midrule
  \mrtwo{tRCD} & \mrtwo \cmdact & \mrtwo{$\rightarrow$} & \mrtwo {\cmdrd/\cmdwr} & Time between an activate
  command and a column command to a bank & \mrtwo{15}\\
  \midrule
  \mrtwo{tRP} & \mrtwo \cmdpre & \mrtwo{$\rightarrow$} & \mrtwo \cmdact & Time taken to complete a precharge
  operation in a bank & \mrtwo{15}\\
  \midrule
  \mrthr{tWR} & \mrthr \cmdwr & \mrthr{$\rightarrow$} & \mrthr \cmdpre & Time taken to ensure that data is
  safely written to the DRAM cells after a write operation (called
  \emph{write recovery}) & \mrthr{15}\\
  \bottomrule
\end{tabular}

  \caption[DDR3-1600 DRAM timing constraints]{Key DRAM timing
    constraints, and their values for DDR3-1600~\cite{ddr3}}
  \label{table:timing-constraints}
\end{table}

\subsection{DRAM Module}
\label{sec:dram-module}

As mentioned before, each \cmdrd or \cmdwr command for a single DRAM
chip typically involves only 64 bits. In order to achieve high memory
bandwidth, commodity DRAM modules group several DRAM chips (typically
4 or 8) together to form a \emph{rank} of DRAM chips. The idea is to
connect all chips of a single rank to the same command and address
buses, while providing each chip with an independent data bus. In
effect, all the chips within a rank receive the same command with the
same address, making the rank a \emph{logically large} DRAM chip.

\begin{figure}[h]
  \centering
  \begin{tikzpicture}[semithick]

  \tikzset{chip/.style={draw,rounded corners=3pt,black!40,fill=black!20,
     minimum width=1.4cm, minimum height=1.75cm,outer sep=0pt,anchor=west}};
  \tikzset{wire/.style={thin,black!80}};
  
  \node (chip0) [chip] at (0,0) {};
  \foreach \x [count=\i] in {0,...,6} {
    \node (chip\i) [chip, xshift=2mm] at (chip\x.east) {};
  }

  \foreach \i in {0,...,7} {
    \coordinate (cmd\i) at ($(chip\i.south) + (-0.3,0)$);
    \coordinate (addr\i) at (chip\i.south);
    \coordinate (data\i) at ($(chip\i.south) + (0.3,0)$);
  }

  \coordinate (cmdorigin) at ($(chip0.south west) + (-1.5,-0.4)$);
  \coordinate (addrorigin) at ($(chip0.south west) + (-1.5,-1)$);
  \draw [wire] (cmdorigin) -- (cmdorigin-|chip7.south east);
  \draw [wire] (addrorigin) -- (addrorigin-|chip7.south east);

  \foreach \i in {0,...,7} {
    \coordinate (dataorg\i) at ($(chip0.south west) + (-1.5,-1.5 - 0.2*\i)$);
    \draw [wire] (dataorg\i) -- (dataorg\i-|chip7.south east);
  }

  \foreach \i in {0,...,7} {
    \node at (chip\i) {Chip \i};
    \draw [wire,fill] (cmd\i) -- (cmd\i|-cmdorigin) circle(1.5pt);
    \draw [wire,fill] (addr\i) -- (addr\i|-addrorigin) circle(1.5pt);
    \draw [wire,fill] (data\i) -- (data\i|-dataorg\i) circle(1.5pt);
  }

  \node at (cmdorigin) [xshift=3mm,fill=white,anchor=west] {\emph{cmd}};
  \node at (addrorigin) [xshift=3mm,fill=white,anchor=west] {\emph{addr}};
  \node at ($(dataorg1)!0.5!(dataorg6)$) [xshift=3mm,fill=white,anchor=west] {\emph{data} (64 bits)};

\end{tikzpicture}
  \caption{Organization of a DRAM rank}
  \label{fig:dram-rank}
\end{figure}
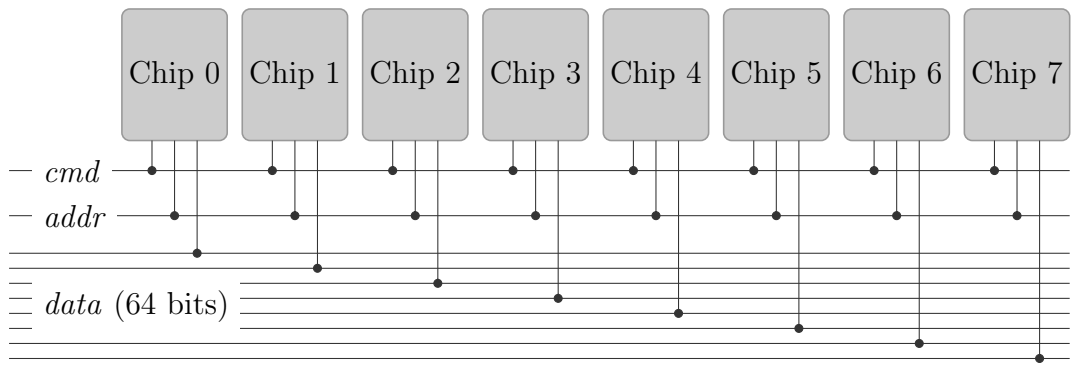

Figure~\ref{fig:dram-rank} shows the logical organization of a DRAM
rank. Many commodity DRAM ranks consist of 8 chips with each chip
accessing 8 bytes of data in response to each \cmdrd or \cmdwr
command. Therefore, in total, each \cmdrd or \cmdwr command accesses
64 bytes of data, the typical cache line size in many processors.

\subsection{RowClone: Bulk Copy and Initialization using DRAM}

RowClone~\cite{rowclone} is a mechanism to perform bulk copy and
initialization operations \emph{completely} inside the DRAM. This approach
obviates the need to transfer large quantities of data on the
memory channel, thereby significantly improving the efficiency of
a bulk copy operation. As bulk data initialization (and specifically
bulk zeroing) can be viewed as a special case of a bulk copy
operation, RowClone can be easily extended to perform such bulk
initialization operations with high efficiency.

\begin{figure}[h]
  \centering
  \includegraphics{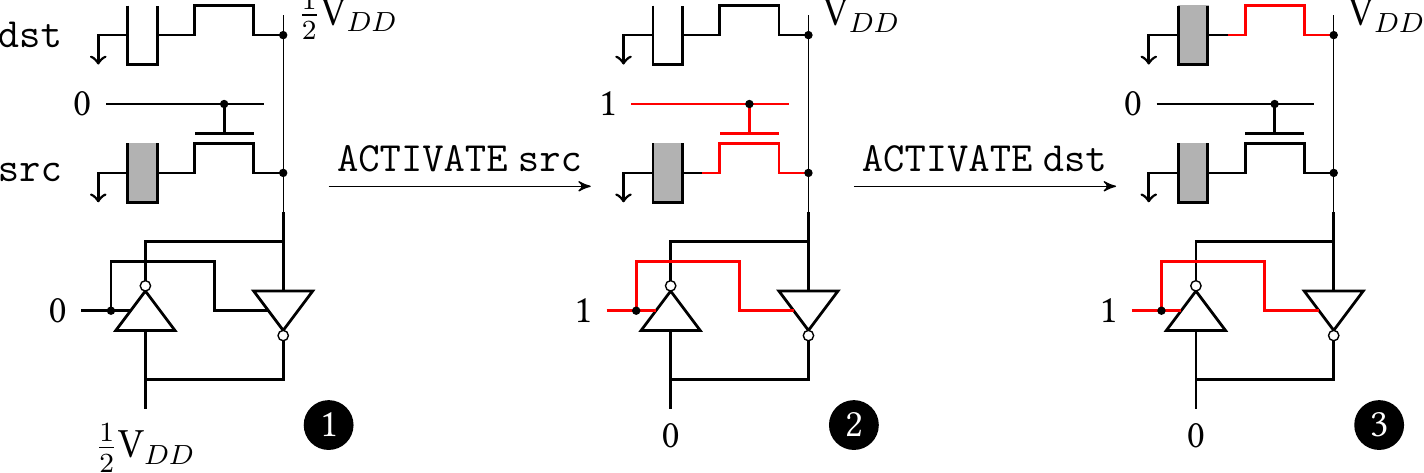}
  \caption{RowClone: Fast Parallel Mode (FPM)}
  \label{fig:cell-fpm}
\end{figure}

RowClone consists of two independent mechanisms that exploit several
observations about DRAM organization and operation.  The first
mechanism, called the \emph{Fast Parallel Mode} (FPM), efficiently
copies data between two rows of DRAM cells that share the same set of
sense amplifiers (i.e., two rows within the same subarray). To copy
data from a source row to a destination row within the same subarray,
RowClone-FPM first issues an \cmdact to the source row, immediately
followed by an \cmdact to the destination row. We show that this
sequence of back-to-back row activations inside the same subarray
results in a data copy from the source row to the destination
row. Figure~\ref{fig:cell-fpm} shows the operation of RowClone-FPM on
a single cell.

Without any further optimizations, the latency of RowClone-FPM is
equivalent to that of two row activations followed by a precharge,
which is an order of magnitude faster than existing
systems~\cite{rowclone}.

\begin{figure}[h]
  \centering
  \includegraphics[angle=90]{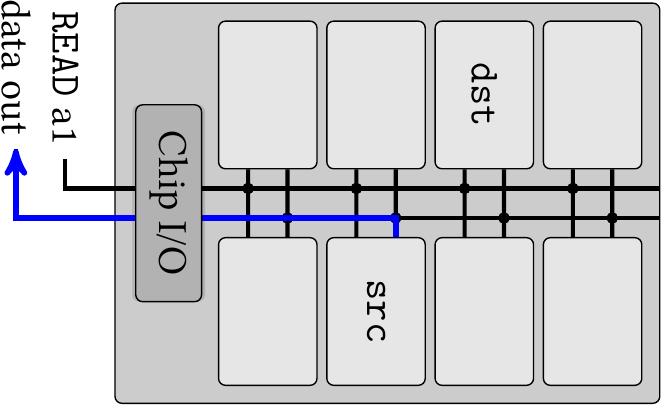}
  \includegraphics[angle=90]{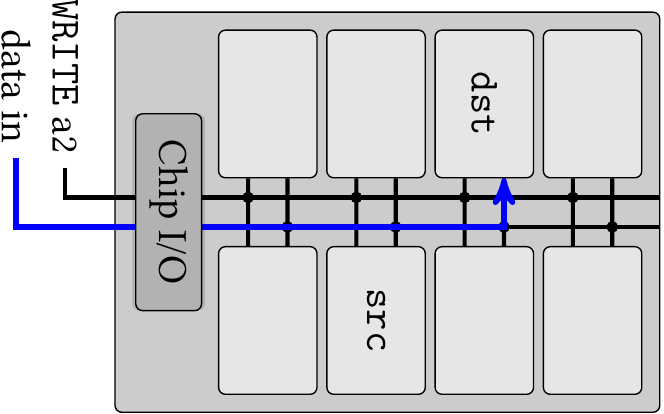}
  \includegraphics[angle=90]{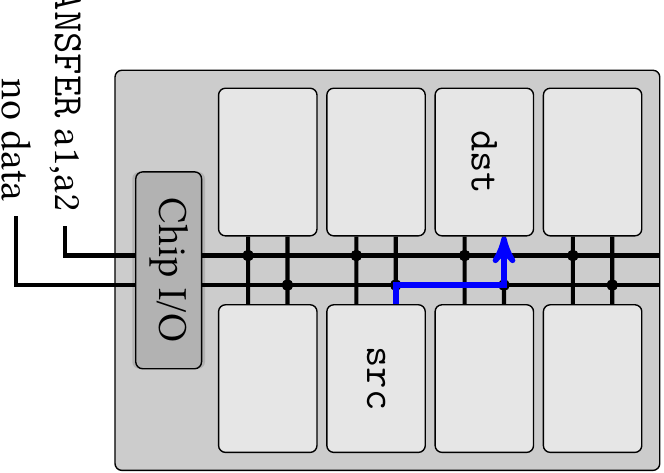}
  \caption{RowClone: Pipelined Serial Mode (PSM)}
  \label{fig:psm}
\end{figure}

The second mechanism, called the \emph{Pipelined Serial Mode} (PSM),
efficiently copies cache lines between two banks within a module in a
pipelined manner. To copy data from a source row in one bank to a
destination row in a different bank, RowClone-PSM first activates both
the rows. It then uses a newly-proposed command, \cmdtr, to copy a
single cache line from the source row \emph{directly} into the
destination row, without having to send the data outside the
chip. Figure~\ref{fig:psm} compares the data path of \cmdrd, \cmdwr,
and \cmdtr. Although not as fast as FPM, PSM has fewer constraints and
hence is more generally applicable~\cite{rowclone}.

We refer the reader to \cite{rowclone} for detailed
information on RowClone.

\section{Ambit: A Bulk Bitwise Execution Engine}
\label{sec:ambit}

In this section, we describe the design and implementation of
Ambit, a new mechanism that converts DRAM into a bulk bitwise
execution engine with low cost. As mentioned in the introduction,
bulk bitwise operations are triggered by many important
applications---e.g., bitmap
indices~\cite{bmide,bmidc,fastbit,oracle,redis,rlite},
databases~\cite{bitweaving,widetable}, document
filtering~\cite{bitfunnel}, DNA
processing~\cite{bitwise-alignment,shd,gatekeeper,grim,myers1999,nanopore-sequencing,shouji}, encryption
algorithms~\cite{xor1,xor2,enc1}, graph
processing~\cite{pinatubo}, and networking~\cite{hacker-delight}.
Accelerating bulk bitwise operations can significantly boost the
performance of these applications.

\subsection{\ambitao}
\label{sec:bitwise-and-or}

The first component of our mechanism, \ambitao, uses the analog
nature of the charge sharing phase to perform bulk bitwise AND and
OR directly inside the DRAM chip. It specifically exploits two
facts about DRAM operation:
\begin{enumerate}[topsep=2pt]\itemsep0pt\parskip0pt
\item In a subarray, each sense amplifier is shared by many (typically
  512 or 1024) DRAM cells on the same bitline.
\item The final state of the bitline after sense amplification
  depends primarily on the voltage deviation on the bitline after
  the charge sharing phase.
\end{enumerate}
Based on these facts, we observe that simultaneously activating three
cells, rather than a single cell, results in a \emph{bitwise majority
  function}---i.e., at least two cells have to be fully charged for
the final state to be a logical ``1''. We refer to simultaneous
activation of three cells (or rows) as \emph{triple-row
  activation}. We now conceptually describe triple-row activation and
how we use it to perform bulk bitwise AND and OR operations.

\subsubsection{Triple-Row Activation (\tra)}
\label{sec:triple-row-activation}

A triple-row activation (\tra) simultaneously connects a sense
amplifier with three DRAM cells on the same bitline. For ease of
conceptual understanding, let us assume that the three cells have the
same capacitance, the transistors and bitlines behave ideally (no
resistance), and the cells start at a fully refreshed state. Then,
based on charge sharing principles~\cite{dram-cd}, the bitline
deviation at the end of the charge sharing phase of the \tra is:
\begingroup\makeatletter\def\f@size{8}\check@mathfonts\vspace{-2mm}
\begin{eqnarray}
  \delta &=& \frac{k.C_c.V_{DD} + C_b.\frac{1}{2}V_{DD}}{3C_c + C_b}
  - \frac{1}{2}V_{DD} \nonumber \\
  &=& \frac{(2k - 3)C_c}{6C_c + 2C_b}V_{DD} \label{eqn:delta}
\end{eqnarray}
\endgroup where, $\delta$ is the bitline deviation, $C_c$ is the cell
capacitance, $C_b$ is the bitline capacitance, and $k$ is the number
of cells in the fully charged state. It is clear that $\delta > 0$ if
and only if $2k - 3 > 0$. In other words, the bitline deviation is
positive if $k = 2, 3$ and it is negative if $k = 0,
1$. Therefore, we expect the final state of the bitline to be \vdd if
at least two of the three cells are initially fully charged, and the
final state to be $0$, if at least two of the three cells are
initially fully empty.

\begin{figure}[h]
  \centering
  \includegraphics[scale=1.4]{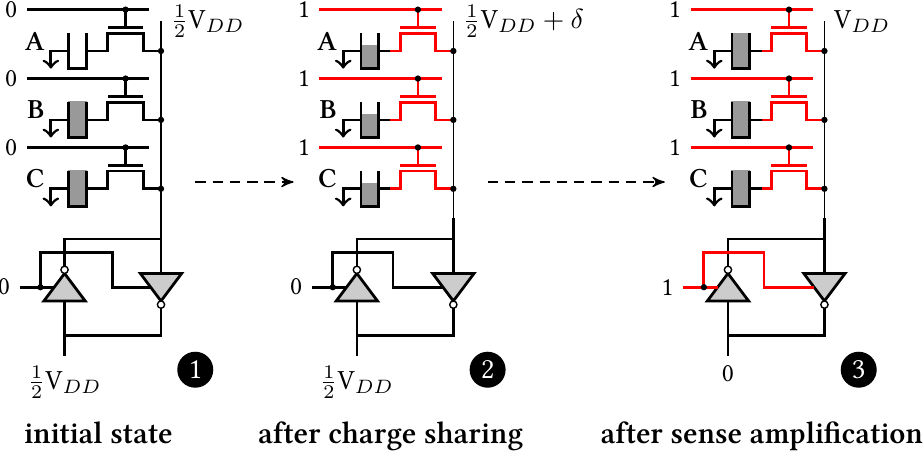}
  \caption{Triple-row activation (TRA)}
  \label{fig:triple-row-activation}
\end{figure}

Figure~\ref{fig:triple-row-activation} shows an example \tra where
two of the three cells are initially in the charged state
\ding{202}. When the wordlines of all the three cells are raised
simultaneously \ding{203}, charge sharing results in a positive
deviation on the bitline. Therefore, after sense amplification
\ding{204}, the sense amplifier drives the bitline to \vdd, and as
a result, fully charges all the three cells.\footnote{Modern DRAMs
  use an open-bitline
  architecture~\cite{diva-dram,lisa,data-retention,dram-cd}
  (Section~\ref{sec:dram-mat}), where cells are also connected to
  \bbar. The three cells in our example are connected to the
  bitline. However, based on the duality principle of Boolean
  algebra~\cite{boolean}, i.e., \bnot(A \band B) $\equiv$ (\bnot
  A) \bor (\bnot B), \tra works seamlessly even if all the three
  cells are connected to \bbar.}

If $A$, $B$, and $C$ represent the logical values of the three cells,
then the final state of the bitline is $AB + BC + CA$ (the
\emph{bitwise majority} function). Importantly, we can rewrite this
expression as $C(A + B) + \overline{C}(AB)$. In other words, by
controlling the value of the cell $C$, we can use \tra to execute a
\emph{bitwise AND} or \emph{bitwise OR} of the cells $A$ and $B$.
Since activation is a row-level operation in DRAM, \tra operates on an
\emph{entire row} of DRAM cells and sense amplifiers, thereby enabling
a multi-kilobyte-wide bitwise AND/OR of two rows.

\subsubsection{Making \tra Work}
\label{sec:and-or-challenges}

There are five potential issues with \tra that we need to resolve for
it to be implementable in a real DRAM design.
\begin{enumerate}[labelindent=0pt,topsep=4pt,leftmargin=*]\itemsep0pt\parskip0pt
\item When simultaneously activating three cells, the deviation on the
  bitline may be smaller than when activating only one cell. This may
  lengthen sense amplification or worse, the sense amplifier may
  detect the wrong value.
\item Equation~\ref{eqn:delta} assumes that all cells have the same
  capacitance, and that the transistors and bitlines behave
  ideally. However, due to process variation, these assumptions are
  not true in real designs~\cite{al-dram,diva-dram,chang2017}. This
  can affect the reliability of \tra, and thus the correctness of its
  results.
\item As shown in Figure~\ref{fig:triple-row-activation}
  (state~\ding{204}), \tra overwrites the data of all the three cells
  with the final result value.  In other words, \tra overwrites all
  source cells, thereby destroying their original values.
  \item Equation~\ref{eqn:delta} assumes that the cells involved in a
    \tra are either fully-charged or fully-empty. However, DRAM cells
    leak charge over time~\cite{raidr}. If the cells involved have
    leaked significantly, \tra may not operate as expected.
  \item Simultaneously activating three \emph{arbitrary} rows inside a
    DRAM subarray requires the memory controller and the row decoder
    to simultaneously communicate and decode three row addresses.
    This introduces a large cost on the address bus and the row
    decoder, potentially tripling these structures, if implemented
    na\"ively.
\end{enumerate}

We address the first two issues by performing rigorous circuit
simulations of \tra. Our results confirm that \tra works as expected
(Section~\ref{sec:spice-sim}).  In Sections~\ref{sec:and-or-mechanism}
and \ref{sec:and-or-rowclone}, we propose a simple implementation of
\ambitao that addresses \emph{all} of the last three issues at low
cost.

\subsubsection{Implementation of \ambitao}
\label{sec:and-or-mechanism}

To solve issues 3, 4, and 5 described in
Section~\ref{sec:and-or-challenges}, our implementation of \ambit
reserves a set of \emph{designated rows} in each subarray that are
used to perform {\tra}s. These designated rows are chosen statically
at \emph{design time}. To perform a bulk bitwise AND or OR operation
on two arbitrary source rows, our mechanism \emph{first copies} the
data of the source rows into the designated rows and performs the
required \tra on the designated rows. As an example, to perform a
bitwise AND/OR of two rows \texttt{A} and \texttt{B}, and store the
result in row \texttt{R}, our mechanism performs the following
steps.\vspace{-1mm}
\begin{enumerate}[topsep=4pt]\itemsep0pt\parsep0pt\parskip0pt\small
\item \emph{Copy} data of row \texttt{A} to designated row \taddr{0}
\item \emph{Copy} data of row \texttt{B} to designated row \taddr{1}
\item \emph{Initialize} designated row \taddr{2} to $0$
\item \emph{Activate} designated rows \taddr{0}, \taddr{1}, and
  \taddr{2} simultaneously
\item \emph{Copy} data of row \taddr{0} to row \texttt{R}
\end{enumerate}\vspace{-1mm}

Let us understand how this implementation addresses the last three
issues described in Section~\ref{sec:and-or-challenges}.  First, by
performing the \tra on the designated rows, and \emph{not} directly on
the source data, our mechanism avoids overwriting the source data
(issue 3). Second, each copy operation refreshes the cells of the
destination row by accessing the row~\cite{raidr}. Also, each copy
operation takes five-six orders of magnitude lower latency
(100~ns---1~${\mu}s$) than the refresh interval (64~ms). Since these
copy operations (Steps 1 and 2 above) are performed \emph{just before}
the \tra, the rows involved in the \tra are very close to the
fully-refreshed state just before the \tra operation (issue
4). Finally, since the \emph{designated} rows are chosen statically at
design time, the \ambit controller uses a reserved address to
communicate the \tra of a \emph{pre-defined} set of three designated
rows. To this end, \ambit reserves a set of row addresses \emph{just}
to trigger {\tra}s. For instance, in our implementation to perform a
\tra of designated rows \taddr{0}, \taddr{1}, and \taddr{2} (Step 4,
above), the \ambit controller simply issues an \cmdact with the
reserved address \baddr{12} (see Section~\ref{sec:address-grouping}
for a full list of reserved addresses). The row decoder maps
\baddr{12} to \emph{all} the three wordlines of the designated rows
\taddr{0}, \taddr{1}, and \taddr{2}. This mechanism requires \emph{no}
changes to the address bus and significantly reduces the cost and
complexity of the row decoder compared to performing \tra on three
\emph{arbitrary} rows (issue 5).

\subsubsection{Fast Row Copy and Initialization Using RowClone}
\label{sec:and-or-rowclone}

Our mechanism needs three row copy operations and one row
initialization operation. These operations, if performed na\"ively, can
nullify the benefits of \ambit, as a row copy or row initialization
performed using the memory controller incurs high
latency~\cite{rowclone,lisa}. Fortunately, a recent work,
RowClone~\cite{rowclone}, proposes two techniques to efficiently copy
data between rows \emph{directly within} DRAM. The first technique,
RowClone-FPM (Fast Parallel Mode), copies data within a subarray by
issuing two back-to-back {\cmdact}s to the source row and the
destination row. This operation takes only 80~ns~\cite{rowclone}. The
second technique, RowClone-PSM (Pipelined Serial Mode), copies data
between two banks by using the internal DRAM bus. Although
RowClone-PSM is faster and more efficient than copying data using the
memory controller, it is significantly slower than RowClone-FPM.

Ambit relies on using RowClone-FPM for most of the copy
operations. To enable this, we propose three ideas. First, to
allow \ambit to perform the initialization operation using
RowClone-FPM, we reserve two \emph{control} rows in each subarray,
\czero and \cone. \czero is initialized to all zeros and \cone is
initialized to all ones. Depending on the operation to be
performed, bitwise AND or OR, \ambit copies the data from \czero
or \cone to the appropriate designated row using
RowClone-FPM. Second, we reserve separate designated rows in
\emph{each} subarray. This allows each subarray to perform bulk
bitwise AND/OR operations on the rows that belong to that subarray
by using RowClone-FPM for all the required copy operations. Third,
to ensure that bulk bitwise operations are predominantly performed
between rows inside the \emph{same} subarray, we rely on 1)~an
accelerator API that allows applications to specify bitvectors
that are likely to be involved in bitwise operations, and 2)~a
driver that maps such bitvectors to the \emph{same} subarray
(described in Section~\ref{sec:ambit-api-driver}). With these
changes, \ambit can use RowClone-FPM for a significant majority of
the bulk copy operations, thereby ensuring high performance for
the bulk bitwise operations.

A recent work, Low-cost Interlinked Subarrays (LISA)~\cite{lisa},
proposes a mechanism to efficiently copy data across subarrays in
the same bank. LISA uses a row of isolation transistors next to
the sense amplifier to control data transfer across two
subarrays. LISA can potentially benefit \ambit by improving the
performance of bulk copy operations. However, as we will describe
in Section~\ref{sec:bitwise-not}, \ambitn also adds transistors
near the sense amplifier, posing some challenges in integrating
LISA and \ambit. Therefore, we leave the exploration of using LISA
to speed up \ambit as part of future work.

\subsection{\ambitn}
\label{sec:bitwise-not}

\begin{figure}[h]
  \centering
  \includegraphics[scale=1]{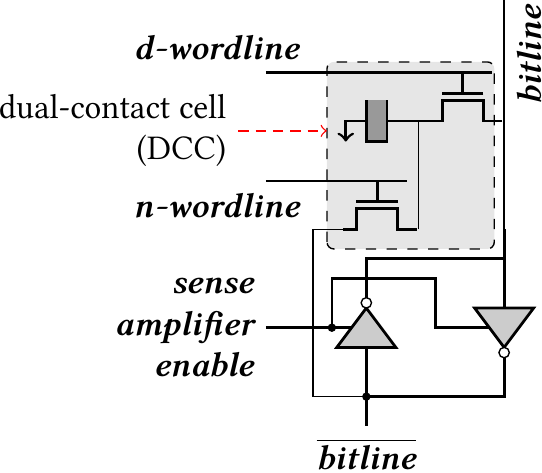}
  \caption{A dual-contact cell connected to a sense
    amplifier}
  \label{fig:dcc-not}
\end{figure}

\ambitn exploits the fact that at the end of the sense amplification
process, the voltage level of the \bbar represents the negated logical
value of the cell.  Our key idea to perform bulk bitwise NOT in DRAM
is to transfer the data on the \bbar to a cell that can \emph{also} be
connected to the bitline. For this purpose, we introduce the
\emph{dual-contact cell} (shown in Figure~\ref{fig:dcc-not}). A
dual-contact cell (DCC) is a DRAM cell with two transistors (a 2T-1C
cell similar to the one described in~\cite{2t-1c-1,migration-cell}).
Figure~\ref{fig:dcc-not} shows a DCC connected to a sense
amplifier. In a DCC, one transistor connects the cell capacitor to the
bitline and the other transistor connects the cell capacitor to the
\bbar.  We refer to the wordline that controls the capacitor-bitline
connection as the \dwordline (or data wordline) and the wordline that
controls the capacitor-\bbar connection as the \nwordline (or negation
wordline).  The layout of the dual-contact cell is similar to Lu et
al.'s migration cell~\cite{migration-cell}.

\begin{figure}[b]
  \centering
  \includegraphics[scale=0.9]{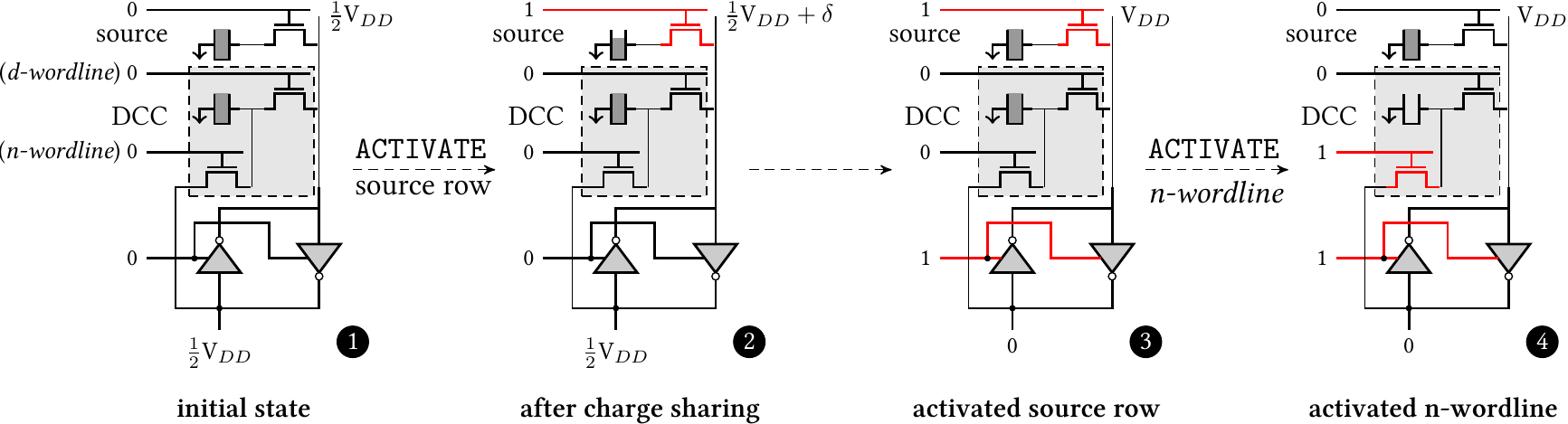}
  \caption{Bitwise NOT using a dual-contact cell}
  \label{fig:bitwise-not}
\end{figure}

Figure~\ref{fig:bitwise-not} shows the steps involved in transferring
the negated value of a \emph{source} cell on to the DCC connected to
the same bitline (i.e., sense amplifier) \ding{202}. Our mechanism
first activates the source cell \ding{203}. The activation drives the
bitline to the data value corresponding to the source cell, \vdd in
this case and the \bbar to the negated value, i.e., 0 \ding{204}. In
this \emph{activated} state, our mechanism activates the
\emph{n-wordline}. Doing so enables the transistor that connects the
DCC to the \bbar~\ding{205}. Since the \bbar is already at a stable
voltage level of $0$, it overwrites the value in the DCC capacitor
with $0$, thereby copying the negated value of the source cell into the
DCC. After this, our mechanism \emph{precharges} the bank, and then
copies the negated value from the DCC to the destination cell using
RowClone.

\textbf{Implementation of \ambitn.}  Based on Lu et
al.'s~\cite{migration-cell} layout, the cost of each row of DCC is
the same as two regular DRAM rows. Similar to the designated rows used for
\ambitao (Section~\ref{sec:and-or-mechanism}), the \ambit controller
uses reserved row addresses to control the \emph{d-wordline}s and
\emph{n-wordline}s of the DCC rows---e.g., in our implementation,
address \baddr{5} maps to the \emph{n-wordline} of the DCC row
(Section~\ref{sec:address-grouping}). To perform a bitwise NOT of row
\texttt{A} and store the result in row \texttt{R}, the \ambit
controller performs the following steps.

\begin{enumerate}[topsep=2pt]\itemsep0pt\parsep0pt\parskip0pt\small
\item \emph{Activate} row \texttt{A}
\item \emph{Activate} \emph{n-wordline} of DCC (address \baddr{5})
\item \emph{Precharge} the bank.
\item \emph{Copy} data from \emph{d-wordline} of DCC to row \texttt{R} (RowClone)
\end{enumerate}\vspace{-1mm}

\section{\ambit: Full Design and Implementation}
\label{sec:implementation}

In this section, we describe our implementation of \ambit by
integrating \ambitao and \ambitn. First, both \ambitao and \ambitn
reserve a set of rows in each subarray and a set of addresses that map
to these rows. We present the full set of reserved addresses and their
mapping in detail (Section~\ref{sec:address-grouping}). Second, we
introduce a new primitive called \aap (\cmdact-\cmdact-\cmdpre) that
the \ambit controller uses to execute various bulk bitwise operations
(Section~\ref{sec:command-sequence}). Third, we describe an
optimization that lowers the latency of the \aap primitive, further
improving the performance of \ambit
(Section~\ref{sec:split-row-decoder}). Fourth, we describe how we
integrate \ambit with the system stack
(Section~\ref{sec:support}). Finally, we evaluate the hardware cost of
\ambit (Section~\ref{sec:hardware-cost}).

\subsection{Row Address Grouping}
\label{sec:address-grouping}

\begin{figure}[h]
  \centering
  \includegraphics[scale=1.3]{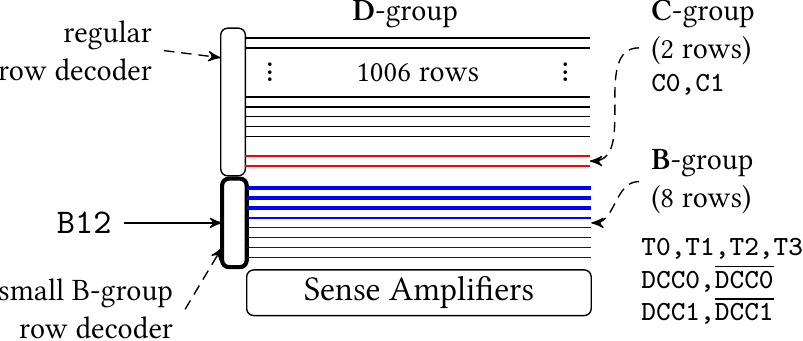}
  \caption{Row address grouping in a subarray. The figure shows how
    the B-group row decoder (Section~\ref{sec:split-row-decoder})
    simultaneously activates rows \taddr{0}, \taddr{1}, and \taddr{2}
    with a single address \baddr{12}.}
  \label{fig:row-address-grouping}
\end{figure}

Our implementation divides the space of row addresses in each
subarray into three distinct groups
(Figure~\ref{fig:row-address-grouping}): 1)~\textbf{B}itwise group,
2)~\textbf{C}ontrol group, and 3)~\textbf{D}ata group.

The \emph{B-group} (or the \emph{bitwise} group) corresponds to the
designated rows used to perform bulk bitwise AND/OR operations
(Section~\ref{sec:and-or-mechanism}) and the dual-contact rows used to
perform bulk bitwise NOT operations
(Section~\ref{sec:bitwise-not}). Minimally, \ambit requires three
designated rows (to perform triple row activations) and one row of
dual-contact cells in each subarray. However, to reduce the number of
copy operations required by certain bitwise operations (like \bxor and
\bxnor), we design each subarray with four designated rows, namely
\taddr{0}---\taddr{3}, and two rows of dual-contact cells, one on each
side of the row of sense amplifiers.\footnote{Each \bxor/\bxnor
  operation involves multiple \band, \bor, and \bnot operations. We
  use the additional designated row and the DCC row to store
  \emph{intermediate} results computed as part of the \bxor/\bxnor
  operation (see Figure~\ref{fig:command-sequences}c).} We refer to
the {\dwordline}s of the two DCC rows as \dccdz and \dccdo, and the
corresponding {\nwordline}s as \dccnz and \dccno. The B-group contains
16 reserved addresses:
\baddr{0}---\baddr{15}. Table~\ref{table:b-group-mapping} lists the
mapping between the 16 addresses and the wordlines. The first eight
addresses individually activate each of the 8 wordlines in the
group. Addresses \baddr{12}---\baddr{15} activate three wordlines
simultaneously. \ambit uses these addresses to trigger triple-row
activations. Finally, addresses \baddr{8}---\baddr{11} activate two
wordlines.  \ambit uses these addresses to copy the result of an
operation simultaneously to two rows. This is useful for \bxor/\bxnor
operations to \emph{simultaneously} negate a row of source data and
also copy the source row to a designated row. Note that this is just
an example implementation of \ambit and a real implementation may use
more designated rows in the B-group, thereby enabling more complex
bulk bitwise operations with fewer copy operations.

\begin{table}[h]
  \centering
  \begin{tabular}{rl}
  \toprule
  \textbf{Addr.} & \textbf{Wordline(s)}\\
  \toprule
  \baddr{0} & \taddr{0}\\
  \baddr{1} & \taddr{1}\\
  \baddr{2} & \taddr{2}\\
  \baddr{3} & \taddr{3}\\
  \baddr{4} & \dccdz\\
  \baddr{5} & \dccnz\\
  \baddr{6} & \dccdo\\
  \baddr{7} & \dccno\\
  \bottomrule
\end{tabular}\quad
\begin{tabular}{rl}
  \toprule
  \textbf{Addr.} & \textbf{Wordline(s)}\\
  \toprule
  \baddr{8} & \dccnz, \taddr{0}\\
  \baddr{9} & \dccno, \taddr{1}\\
  \baddr{10} & \taddr{2}, \taddr{3}\\
  \baddr{11} & \taddr{0}, \taddr{3}\\
  \baddr{12} & \taddr{0}, \taddr{1}, \taddr{2}\\
  \baddr{13} & \taddr{1}, \taddr{2}, \taddr{3}\\
  \baddr{14} & \dccdz, \taddr{1}, \taddr{2}\\
  \baddr{15} & \dccdo, \taddr{0}, \taddr{3}\\
  \bottomrule
\end{tabular}

  \caption{Mapping of B-group addresses to corresponding activated
    wordlines}
  \label{table:b-group-mapping}
\end{table}

The \emph{C-group} (or the \emph{control} group) contains the two
pre-initialized rows for controlling the bitwise AND/OR operations
(Section~\ref{sec:and-or-rowclone}). Specifically, this group contains
two addresses: \czero (row with all zeros) and \cone (row with all
ones).

The \emph{D-group} (or the \emph{data} group) corresponds to the rows
that store regular data. This group contains all the addresses that
are neither in the \emph{B-group} nor in the
\emph{C-group}. Specifically, if each subarray contains $1024$ rows,
then the \emph{D-group} contains $1006$ addresses, labeled
\daddr{0}---\daddr{1005}. \ambit exposes only the D-group addresses to
the software stack. To ensure that the software stack has a contiguous
view of memory, the \ambit controller interleaves the row addresses
such that the D-group addresses across all subarrays are mapped
contiguously to the processor's physical address space.

With these groups, the \ambit controller can use the \emph{existing}
DRAM interface to communicate all variants of \cmdact to the \ambit
chip \emph{without} requiring new commands. Depending on the address
group, the \ambit DRAM chip internally processes each \cmdact
appropriately.  For instance, by just issuing an \cmdact to address
\baddr{12}, the \ambit controller triggers a triple-row activation of
\taddr{0}, \taddr{1}, and \taddr{2}. We now describe how the \ambit
controller uses this row address mapping to perform bulk bitwise
operations.

\subsection{Executing Bitwise Ops: The AAP Primitive}
\label{sec:command-sequence}
Let us consider the operation, \daddr{k} \texttt{=} \bnot
\daddr{i}. To perform this bitwise-NOT operation, the \ambit
controller sends the following sequence of commands.

{\tabcolsep4pt\small
\begin{tabular}{llllll}
1. & \cmdact \daddr{i}; & 2. & \cmdact \baddr{5}; & 3. & \cmdpre;\\
4. & \cmdact \baddr{4}; & 5. & \cmdact \daddr{k}; & 6. & \cmdpre;
\end{tabular}
}

The first three steps are the same as those described in
Section~\ref{sec:bitwise-not}. These three operations copy the negated
value of row \daddr{i} into the \dccdz row (as described in
Figure~\ref{fig:bitwise-not}). Step 4 activates \dccdz, the \dwordline
of the first DCC row, transferring the negated source data onto the
bitlines. Step 5 activates the destination row, copying the data on
the bitlines, i.e., the negated source data, to the destination
row. Step 6 prepares the array for the next access by issuing a
\cmdpre.

The bitwise-NOT operation consists of two steps of
\cmdact-\cmdact-\cmdpre operations. We refer to this sequence as the
\aap primitive. Each \aap takes two addresses as
input. \texttt{\aap(addr1, addr2)} corresponds to the following
sequence of commands:\vspace{3pt}\\\vspace{3pt}
\centerline{\texttt{\cmdact addr1; \cmdact addr2; \cmdpre;}}
Logically, an \aap operation copies the result of the row activation
of the first address (\texttt{addr1}) to the row(s) mapped to the
second address (\texttt{addr2}).

\begin{figure}[h]
  \centering
  \includegraphics[scale=1.3]{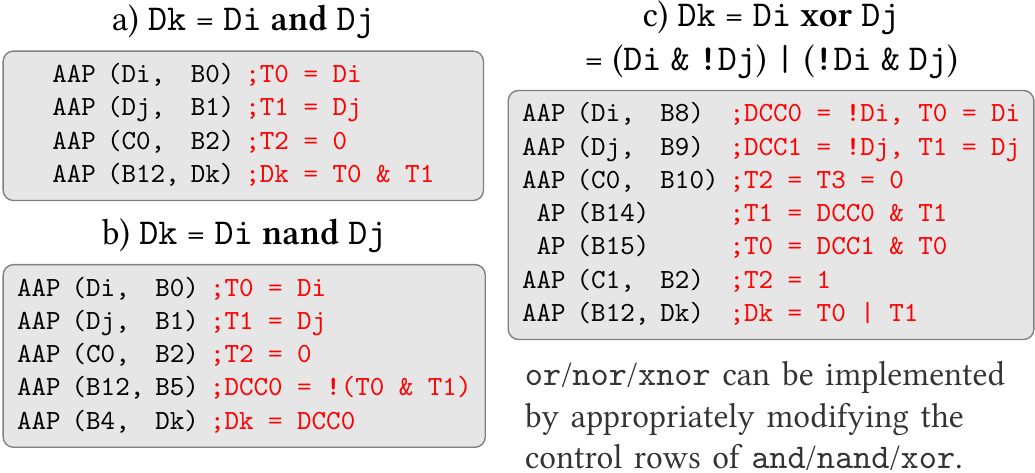}
  \caption{Command sequences for different bitwise operations}
  \label{fig:command-sequences}
\end{figure}

Most bulk bitwise operations mainly involve a sequence of \aap
operations. In a few cases, they require a regular \cmdact followed by
a \cmdpre, which we refer to as \ap.  \ap takes one address as
input. \texttt{\ap(addr)} maps to the following two
commands:\vspace{3pt}\\\vspace{3pt} \centerline{\texttt{\cmdact addr;
    \cmdpre;}} Figure~\ref{fig:command-sequences} shows the steps
taken by the \ambit controller to execute three bulk bitwise
operations: \band, \bnand, and \bxor.

Let us consider the \band operation, \daddr{k} = \daddr{i} \band
\daddr{j}, shown in Figure~\ref{fig:command-sequences}a. The four
\aap operations directly map to the steps described in
Section~\ref{sec:and-or-mechanism}. The first \aap copies the first
source row (\daddr{i}) into the designated row \taddr{0}. Similarly,
the second \aap copies the second source row \daddr{j} to row
\taddr{1}, and the third \aap copies the control row ``0'' to row
\taddr{2} (to perform a bulk bitwise AND). Finally, the last \aap
1)~issues an \cmdact to address \baddr{12}, which simultaneously
activates the rows \taddr{0}, \taddr{1}, and \taddr{2}, resulting in
an \band operation of the rows \taddr{0} and \taddr{1}, 2) issues an
\cmdact to \daddr{k}, which copies the result of the \band operation
to the destination row \daddr{k}, and 3)~precharges the bank to
prepare it for the next access.

While each bulk bitwise operation involves multiple copy operations,
this copy overhead can be reduced by applying standard compilation
techniques. For instance, accumulation-like operations generate
intermediate results that are immediately consumed. An optimization
like dead-store elimination may prevent these values from being copied
unnecessarily. Our evaluations (Section~\ref{sec:applications}) take
into account the overhead of the copy operations \emph{without} such
optimizations.

\subsection{Accelerating AAP with a Split Row Decoder}
\label{sec:split-row-decoder}

The latency of executing any bulk bitwise operation using \ambit
depends on the latency of the \aap primitive. The latency of the \aap
in turn depends on the latency of \cmdact, i.e., \tras, and the
latency of \cmdpre, i.e., \trp. The na\"ive approach to execute an \aap
is to perform the three operations serially. Using this approach, the
latency of \aap is 2\tras + \trp (80~ns for
DDR3-1600~\cite{ddr3-1600}). While even this na\"ive approach offers
better throughput and energy efficiency than existing systems (not
shown here), we propose a simple optimization that significantly
reduces the latency of \aap.

Our optimization is based on two observations. First, the second
\cmdact of an \aap is issued to an already \emph{activated bank}. As a
result, this \cmdact does \emph{not} require \emph{full} sense
amplification, which is the dominant portion of
\tras~\cite{diva-dram,al-dram,chargecache}. This enables the
opportunity to reduce the latency for the second \cmdact of each
\aap. Second, when we examine all the bitwise operations in
Figure~\ref{fig:command-sequences}, with the exception of one \aap in
\bnand, we find that \emph{exactly one} of the two \cmdacts in each
\aap is to a \emph{B-group} address. This enables the opportunity to
use a \emph{separate} decoder for \emph{B-group} addresses, thereby
overlapping the latency of the two row activations in each \aap.

To exploit both of these observations, our mechanism splits the row
decoder into two parts. The first part decodes all \emph{C/D-group}
addresses and the second smaller part decodes \emph{only}
\emph{B-group} addresses. Such a split allows the subarray to
\emph{simultaneously} decode a \emph{C/D-group} address along with a
\emph{B-group} address. When executing an \aap, the \ambit controller
issues the second \cmdact of an \aap after the first activation has
sufficiently progressed. This forces the sense amplifier to overwrite
the data of the second row to the result of the first activation. This
mechanism allows the \ambit controller to significantly overlap the
latency of the two \cmdacts. This approach is similar to the
inter-segment copy operation used by Tiered-Latency
DRAM~\cite{tl-dram}. Based on SPICE simulations, our estimate of the
latency of executing the back-to-back {\cmdact}s is only 4~ns larger
than \tras. For DDR3-1600 (8-8-8) timing parameters~\cite{ddr3-1600},
this optimization reduces the latency of \aap from 80~ns to 49~ns.

Since only addresses in the \emph{B-group} are involved in triple-row
activations, the complexity of simultaneously raising three wordlines
is restricted to the small \emph{B-group} decoder. As a result, the
split row decoder also reduces the complexity of the changes \ambit
introduces to the row decoding logic.

\section{Integrating \ambit with the System}
\label{sec:support}

\ambit can be plugged in as an I/O (e.g., PCIe) device and interfaced
with the CPU using a device model similar to other accelerators (e.g.,
GPU). While this approach is simple, as described in previous
sections, the address and command interface of \ambit is \emph{exactly
  the same} as that of commodity DRAM. This enables the opportunity to
directly plug \ambit onto the system memory bus and control it using
the memory controller. This approach has several benefits. First,
applications can \emph{directly} trigger \ambit operations using CPU
instructions rather than going through a device API, which incurs
additional overhead. Second, since the CPU can \emph{directly} access
\ambit memory, there is no need to copy data between the CPU memory
and the accelerator memory. Third, existing cache coherence protocols
can be used to keep \ambit memory and the on-chip cache coherent. To
plug \ambit onto the system memory bus, we need additional support
from the rest of the system stack, which we describe in this section.

\subsection{ISA Support}

To enable applications to communicate occurrences of bulk bitwise
operations to the processor, we introduce new instructions of the
form,\\ \centerline{\texttt{bbop dst, src1, [src2], size}}

where \texttt{bbop} is the bulk bitwise operation, \texttt{dst} is the
destination address, \texttt{src1} and \texttt{src2} are the source
addresses, and \texttt{size} denotes the length of operation in
bytes. Note that \texttt{size} must be a multiple of DRAM row
size. For bitvectors that are not a multiple of DRAM row size, we
assume that the application will appropriately pad them with dummy
data, or perform the residual (sub-row-sized) operations using the
CPU.

\subsection{\ambit API/Driver Support}
\label{sec:ambit-api-driver}

For \ambit to provide significant performance benefit over existing
systems, it is critical to ensure that most of the required copy
operations are performed using RowClone-FPM, i.e., the source rows and
the destination rows involved in bulk bitwise operations are present
in the same DRAM subarray. To this end, we expect the manufacturer of
\ambit to provide 1)~an API that enables applications to specify
bitvectors that are likely to be involved in bitwise operations, and
2)~a driver that is aware of the internal mapping of DRAM rows to
subarrays and maps the bitvectors involved in bulk bitwise operations
to the same DRAM subarray. Note that for a large bitvector, \ambit
does \emph{not} require the entire bitvector to fit inside a
\emph{single} subarray. Rather, each bitvector can be interleaved
across multiple subarrays such that the corresponding portions of each
bitvector are in the same subarray. Since each subarray contains over
1000 rows to store application data, an application can map hundreds
of \emph{large} bitvectors to \ambit, such that the copy operations
required by \emph{all} the bitwise operations across all these
bitvectors can be performed efficiently using RowClone-FPM.

\subsection{Implementing the \texttt{bbop} Instructions }

Since all \ambit operations are row-wide, \ambit requires the source
and destination rows to be row-aligned and the size of the operation
to be a multiple of the size of a DRAM row. The microarchitecture
implementation of a \texttt{bbop} instruction checks if each instance
of the instruction satisfies this constraint. If so, the CPU sends the
operation to the memory controller, which completes the operation
using \ambit. Otherwise, the CPU executes the operation itself.

\subsection{Maintaining On-chip Cache Coherence }

Since both CPU and \ambit can access/modify data in memory, before
performing any \ambit operation, the memory controller must 1)~flush
any dirty cache lines from the source rows, and 2)~invalidate any
cache lines from destination rows. Such a mechanism is already
required by Direct Memory Access (DMA)~\cite{linux-dma}, which is
supported by most modern processors, and also by recently proposed
mechanisms~\cite{rowclone,pointer3d,lazypim-cal,lazypim-arxiv}. As \ambit operations are always
row-wide, we can use structures like the Dirty-Block Index~\cite{dbi}
to speed up flushing dirty data. Our mechanism invalidates the cache
lines of the destination rows in parallel with the \ambit
operation. Other recently-proposed techniques like
LazyPIM~\cite{lazypim-cal} and CoNDA~\cite{conda} can also be
employed with Ambit.

\subsection{Error Correction and Data Scrambling}
\label{sec:ecc}

In DRAM modules that support Error Correction Code (ECC), the memory
controller must first read the data and ECC to verify data
integrity. Since Ambit reads and modifies data directly in memory, it
does not work with the existing ECC schemes (e.g.,
SECDED~\cite{secded}). To support ECC with Ambit, we need an ECC
scheme that is homomorphic~\cite{homomorphism} over all bitwise
operations, i.e., ECC(A \band B) = ECC(A) \band ECC(B), and similarly
for other bitwise operations. The only scheme that we are aware of
that has this property is triple modular redundancy (TMR)~\cite{tmr},
wherein ECC(A) = AA. The design of lower-overhead ECC schemes that are
homomorphic over all bitwise operations is an open problem. For the
same reason, \ambit does \emph{not} work with data scrambling
mechanisms that pseudo-randomly modify the data written to
DRAM~\cite{data-scrambling}. Note that these challenges are also
present in any mechanism that \emph{interprets} data directly in
memory (e.g., the Automata Processor~\cite{automata,pap}). We leave
the evaluation of Ambit with TMR and exploration of other ECC and data
scrambling schemes to future work.

\subsection{\ambit Hardware Cost}
\label{sec:hardware-cost}

As \ambit largely exploits the structure and operation of existing
DRAM design, we estimate its hardware cost in terms of the overhead it
imposes on top of today's DRAM chip and memory controller.

\subsubsection{\ambit Chip Cost}

In addition to support for RowClone, \ambit has only two changes on
top of the existing DRAM chip design. First, it requires the row
decoding logic to distinguish between the \emph{B-group} addresses and
the remaining addresses. Within the \emph{B-group}, it must implement
the mapping described in Table~\ref{table:b-group-mapping}. As the
\emph{B-group} contains only 16 addresses, the complexity of the
changes to the row decoding logic is low.  The second source of cost
is the implementation of the dual-contact cells (DCCs). In our design,
each sense amplifier has only one DCC on each side, and each DCC has
two wordlines associated with it. In terms of area, each DCC row costs
roughly two DRAM rows, based on estimates from Lu et
al.~\cite{migration-cell}. We estimate the overall storage cost of
\ambit to be roughly 8 DRAM rows per subarray---for the four
designated rows and the DCC rows ($< 1\%$ of DRAM chip area).

\subsubsection{\ambit Controller Cost}

On top of the existing memory controller, the \ambit controller must
statically store 1)~information about different address groups, 2)~the
timing of different variants of the \cmdact, and 3)~the sequence of
commands required to complete different bitwise operations. When
\ambit is plugged onto the system memory bus, the controller can
interleave the various AAP operations in the bitwise operations with
other regular memory requests from different applications. For this
purpose, the \ambit controller must also track the status of on-going
bitwise operations. We expect the overhead of these additional pieces
of information to be small compared to the benefits enabled by \ambit.

\subsubsection{\ambit Testing Cost}
\label{sec:testing-cost}

Testing \ambit chips is similar to testing regular DRAM chips. In
addition to the regular DRAM rows, the manufacturer must test if the
\tra operations and the DCC rows work as expected. In each subarray
with 1024 rows, these operations concern only 8 DRAM rows and 16
addresses of the B-group. In addition, all these operations are
triggered using the \cmdact command. Therefore, we expect the overhead
of testing an \ambit chip on top of testing a regular DRAM chip to be
low.

When a component is found to be faulty during testing, DRAM
manufacturers use a number of techniques to improve the overall
yield; The most prominent among them is using spare rows to
replace faulty DRAM rows. Similar to some prior
works~\cite{rowclone,tl-dram,al-dram,salp}, \ambit requires faulty
rows to be mapped to spare rows within the \emph{same}
subarray. Note that, since \ambit preserves the existing DRAM
command interface, an \ambit chip that fails during testing can
still be shipped as a regular DRAM chip. This significantly
reduces any potential negative impact of \ambit-specific failures
on overall DRAM yield.

\section{Circuit-level SPICE Simulations}
\label{sec:spice-sim}

\newcommand{\pvar}[1]{$\pm$#1\%}

We use SPICE simulations to confirm that \ambit works reliably. Of the
two components of \ambit, our SPICE results show that \ambitn
\emph{always} works as expected and is \emph{not} affected by process
variation. This is because, \ambitn operation is very similar to
existing DRAM operation (Section~\ref{sec:bitwise-not}).  On the other
hand, \ambitao requires triple-row activation, which involves charge
sharing between three cells on a bitline. As a result, it can be
affected by process variation in various circuit components.

To study the effect of process variation on \tra, our SPICE
simulations model variation in \emph{all} the components in the
subarray (cell capacitance; transistor length, width, resistance;
bitline/wordline capacitance and resistance; voltage levels).  We
implement the sense amplifier using 55nm DDR3 model
parameters~\cite{rambus}, and PTM low-power transistor
models~\cite{ptmweb,zhaoptm}. We use cell/transistor parameters
from the Rambus power model~\cite{rambus} (cell capacitance =
22fF; transistor width/height = 55nm/85nm).\footnote{In DRAM,
  temperature affects mainly cell
  leakage~\cite{al-dram,data-retention,vrt-1,vrt-2,raidr,reaper,chang2017,softmc,rowhammer,dram-latency-puf,fly-dram,vampire,drange}. As
  \tra is performed on cells that are almost fully-refreshed, we
  do not expect temperature to affect TRA.}
%% REVISION

We first identify the worst case for \tra, wherein every component has
process variation that works toward making \tra fail. Our results show
that even in this extremely adversarial scenario, \tra works reliably
for up to \pvar{6} variation in each component.

In practice, variations across components are not so highly
correlated. Therefore, we use Monte-Carlo simulations to understand
the practical impact of process variation on \tra. We increase the
amount of process variation from \pvar{5} to \pvar{25} and run 100,000
simulations for each level of process
variation. Table~\ref{tab:mc-spice} shows the percentage of iterations
in which \tra operates incorrectly for each level of variation.

\begin{table}[h]
  \centering
  \begin{tabular}{rcccccc}
  \toprule
  Variation & \pvar{0} & \pvar{5} & \pvar{10} & \pvar{15} &
  \pvar{20} & \pvar{25}\\
  \midrule
  \% Failures & 0.00\% & 0.00\% & 0.29\% & 6.01\% & 16.36\% & 26.19\% \\
  \bottomrule
\end{tabular}

  \caption{Effect of process variation on \tra}
  \label{tab:mc-spice}
\end{table}

Two conclusions are in order. First, as expected, up to \pvar{5}
variation, there are zero errors in \tra. Second, even with \pvar{10}
and \pvar{15} variation, the percentage of erroneous {\tra}s across
100,000 iterations each is just 0.29\% and 6.01\%. These results show
that \ambit is reliable even in the presence of significant process
variation.

The effect of process variation is expected to get worse with smaller
technology nodes~\cite{kang2014}. However, as Ambit largely uses the
existing DRAM structure and operation, many techniques used to combat
process variation in existing chips can be used for Ambit as well
(e.g., spare rows or columns). In addition, as described in
Section~\ref{sec:testing-cost}, Ambit chips that fail testing only for
\tra can potentially be shipped as regular DRAM chips, thereby
alleviating the impact of \tra failures on overall DRAM yield, and
thus cost.

Note that if an application can tolerate errors in computation, as
observed by prior
works~\cite{approx-computing,yixin-dsn,rfvp-taco,rfvp-pact}, then \ambit can
be used as an approximate in-DRAM computation substrate even in
the presence of significant process variation. Future work can
potentially explore such an approximate \ambit substrate.

\section{Analysis of Ambit's Throughput \& Energy}
\label{sec:lte-analysis}

\begin{figure}[h]
  \centering
  \includegraphics[scale=1.5]{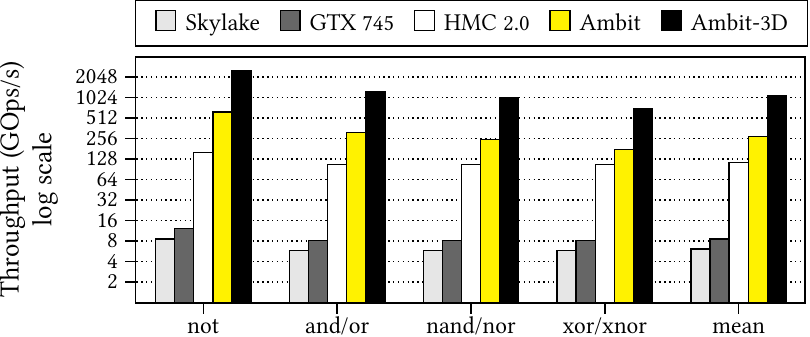}
  \caption{Throughput of bulk bitwise operations.}
  \label{plot:cgb-throughput}
\end{figure}

We compare the raw throughput of bulk bitwise operations using \ambit
to a multi-core Intel Skylake CPU~\cite{intel-skylake}, an NVIDIA
GeForce GTX 745 GPU~\cite{gtx745}, and processing in the logic layer
of an HMC 2.0~\cite{hmc2} device. The Intel CPU has 4 cores with
Advanced Vector eXtensions~\cite{intel-avx}, and two 64-bit DDR3-2133
channels. The GTX 745 contains 3 streaming multi-processors, each with
128 CUDA cores~\cite{tesla}, and one 128-bit DDR3-1800 channel. The
HMC 2.0 device consists of 32 vaults each with 10 GB/s bandwidth. We
use two \ambit configurations: \emph{\ambit} that integrates our
mechanism into a regular DRAM module with 8 banks, and
\emph{\ambit-3D} that extends a 3D-stacked DRAM similar to HMC with
support for \ambit.  For each bitwise operation, we run a
microbenchmark that performs the operation repeatedly for many
iterations on large input vectors (32~MB), and measure the throughput
of the operation.  Figure~\ref{plot:cgb-throughput} plots the results
of this experiment for the five systems (the y-axis is in log scale).

We draw three conclusions. First, the throughput of Skylake, GTX 745,
and HMC 2.0 are limited by the memory bandwidth available to the
respective processors. With an order of magnitude higher available
memory bandwidth, HMC 2.0 achieves 18.5X and 13.1X better throughput
for bulk bitwise operations compared to Skylake and GTX 745,
respectively. Second, \ambit, with its ability to exploit the maximum
internal DRAM bandwidth and memory-level parallelism, outperforms all
three systems. On average, \ambit (with 8 DRAM banks) outperforms
Skylake by 44.9X, GTX 745 by 32.0X, and HMC 2.0 by 2.4X. Third,
3D-stacked DRAM architectures like HMC contain a large number of banks
(256 banks in 4GB HMC 2.0). By extending 3D-stacked DRAM with support
for \ambit, \ambit-3D improves the throughput of bulk bitwise
operations by 9.7X compared to HMC 2.0.

We estimate energy for DDR3-1333 using the Rambus power
model~\cite{rambus}. Our energy numbers include only the DRAM and
channel energy, and not the energy consumed by the processor. For
\ambit, some activations have to raise multiple wordlines and hence,
consume higher energy. Based on our analysis, the activation energy
increases by 22\% for each additional wordline raised.
Table~\ref{table:energy} shows the energy consumed per kilo-byte for
different bitwise operations.  Across all bitwise operations, \ambit
reduces energy consumption by 25.1X---59.5X compared to copying data
with the memory controller using the DDR3 interface.

\begin{table}[h]
  \centering
  \setlength{\tabcolsep}{5pt}
\begin{tabular}{ccrrrr}
  \toprule
   & Design & \bnot & \band/\bor & \bnand/\bnor & \bxor/\bxnor\\
  \toprule
   \multirow{1}{*}{DRAM \&} & \bf DDR3 & 93.7 & 137.9 & 137.9 & 137.9\\
   \multirow{1}{*}{Channel Energy} & \bf \ambit & 1.6 & 3.2 & 4.0 & 5.5\\
  \multirow{1}{*}{(nJ/KB)} &  ($\downarrow$) & 59.5X & 43.9X & 35.1X & 25.1X \\
  \bottomrule  
\end{tabular}

  \caption{Energy of bitwise operations. ($\downarrow$) indicates
    energy reduction  of \ambit over the traditional DDR3-based design.}
  \label{table:energy}
\end{table}

\section{Effect on Real-World Applications}
\label{sec:applications}

We evaluate the benefits of \ambit on real-world applications using
the Gem5 full-system simulator~\cite{gem5}. Table~\ref{tab:parameters}
lists the main simulation parameters. Our simulations take into
account the cost of maintaining coherence, and the overhead of
RowClone to perform copy operations. We assume that application data
is mapped such that all bitwise operations happen across rows
\emph{within a subarray}. We quantitatively evaluate three
applications: 1)~a database bitmap
index~\cite{oracle,redis,rlite,fastbit},
2)~BitWeaving~\cite{bitweaving}, a mechanism to accelerate database
column scan operations, and 3)~a bitvector-based implementation of the
widely-used \emph{set} data structure.  In
Section~\ref{sec:other-apps}, we discuss four other applications that
can benefit from \ambit.

\begin{table}[h]
  \centering
  \begin{tabular}{ll}
    \toprule
    \multirow{2}{*}{Processor} & x86, 8-wide, out-of-order, 4~Ghz\\
    & 64-entry  instruction queue\\
    \midrule
    L1 cache & 32~KB D-cache, 32~KB I-cache, LRU policy\\
    \midrule
    L2 cache & 2~MB, LRU policy, 64~B cache line size\\
    \midrule
    Memory Controller & 8~KB row size,
    FR-FCFS~\cite{frfcfs,frfcfs-patent} scheduling\\
    \midrule
    Main memory & DDR4-2400, 1-channel, 1-rank, 16 banks\\
    \bottomrule
  \end{tabular}
  \caption{Major simulation parameters}
  \label{tab:parameters}
\end{table}

\subsection{Bitmap Indices}
\label{sec:bitmap-indices}

Bitmap indices~\cite{bmide} are an alternative to traditional B-tree
indices for databases. Compared to B-trees, bitmap indices 1)~consume
less space, and 2)~can perform better for many queries (e.g., joins,
scans). Several major databases support bitmap indices
(e.g.,~Oracle~\cite{oracle}, Redis~\cite{redis},
Fastbit~\cite{fastbit}, rlite~\cite{rlite}). Several real applications
(e.g.,~\cite{spool,belly,bitmapist,ai}) use bitmap indices for fast
analytics.  As bitmap indices heavily rely on bulk bitwise operations,
\ambit can accelerate bitmap indices, thereby improving overall
application performance.

\begin{figure}[h]
  \centering
  \includegraphics[scale=1.5]{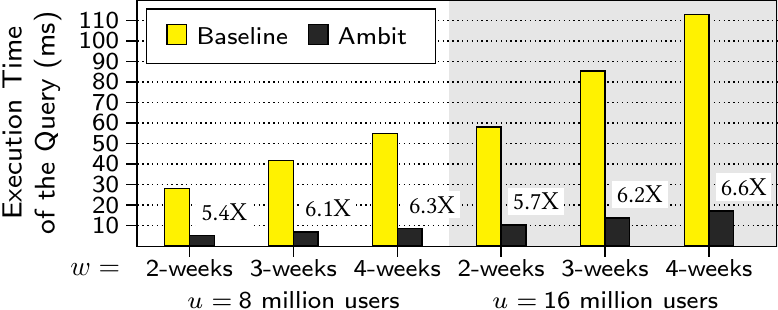}
  \caption{Bitmap index performance. The value above each bar
    indicates the reduction in execution time due to \ambit.}
  \label{fig:rlite}
\end{figure}

To demonstrate this benefit, we use the following workload from a real
application~\cite{ai}. The application uses bitmap indices to track
users' characteristics (e.g., gender) and activities (e.g., did the
user log in to the website on day 'X'?)  for $u$ users.  Our workload
runs the following query: ``How many unique users were active every
week for the past $w$ weeks? and How many male users were active each
of the past $w$ weeks?''  Executing this query requires 6$w$ bulk
bitwise \bor, 2$w$-1 bulk bitwise \band, and $w$+1 bulk bitcount
operations. In our mechanism, the bitcount operations are performed by
the CPU.  Figure~\ref{fig:rlite} shows the end-to-end query execution
time of the baseline and \ambit for the above experiment for various
values of $u$ and $w$.

We draw two conclusions. First, as each query has $O(w)$ bulk bitwise
operations and each bulk bitwise operation takes $O(u)$ time, the
query execution time increases with increasing value $uw$. Second,
\ambit significantly reduces the query execution time compared to the
baseline, by 6X on average.

While we demonstrate the benefits of \ambit using one query, as all
bitmap index queries involve several bulk bitwise operations, we
expect \ambit to provide similar performance benefits for any
application using bitmap indices.

\subsection{BitWeaving: Fast Scans using Bitwise Operations}
\label{sec:bitweaving}

Column scan operations are a common part of many database
queries. They are typically performed as part of evaluating a
predicate. For a column with integer values, a predicate is typically
of the form, \texttt{c1 <= val <= c2}, for two integer constants
\texttt{c1} and \texttt{c2}. Recent
works~\cite{bitweaving,vectorizing-column-scans} observe that existing
data representations for storing columnar data are inefficient for
such predicate evaluation especially when the number of bits used to
store each value of the column is less than the processor word size
(typically 32 or 64). This is because 1)~the values do not align well
with word boundaries, and 2)~the processor typically does not have
comparison instructions at granularities smaller than the word
size. To address this problem, BitWeaving~\cite{bitweaving} proposes
two column representations, called BitWeaving-H and BitWeaving-V. As
BitWeaving-V is faster than BitWeaving-H, we focus our attention on
BitWeaving-V, and refer to it as just BitWeaving.

BitWeaving stores the values of a column such that the first bit of
all the values of the column are stored contiguously, the second bit
of all the values of the column are stored contiguously, and so
on. Using this representation, the predicate \texttt{c1 <= val <= c2},
can be represented as a series of bitwise operations starting from the
most significant bit all the way to the least significant bit (we
refer the reader to the BitWeaving paper~\cite{bitweaving} for the
detailed algorithm). As these bitwise operations can be performed in
parallel across multiple values of the column, BitWeaving uses the
hardware SIMD support to accelerate these operations. With support for
\ambit, these operations can be performed in parallel across a larger
set of values compared to 128/256-bit SIMD available in existing CPUs,
thereby enabling higher performance.

We show this benefit by comparing the performance of BitWeaving using
a baseline CPU with support for 128-bit SIMD to the performance of
BitWeaving accelerated by \ambit for the following commonly-used query
on a table \texttt{T}:\\ \centerline{\small{`\texttt{select count(*)
      from T where c1 <= val <= c2}'}}

\begin{figure}[h]
  \centering
  \includegraphics[scale=1.5]{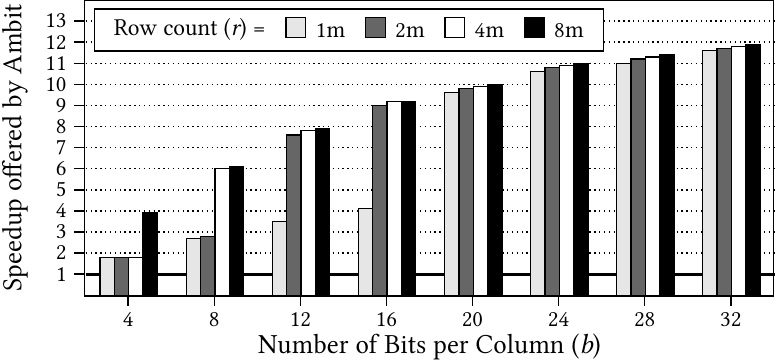}
  \caption{Speedup offered by \ambit for BitWeaving over our baseline
    CPU with SIMD support}
  \label{fig:bitweaving}
\end{figure}

Evaluating the predicate involves a series of bulk bitwise operations
and the \texttt{count(*)} requires a bitcount operation. The execution
time of the query depends on 1)~the number of bits (\emph{b}) used to
represent each value \texttt{val}, and 2)~the number of rows
(\emph{r}) in the table \texttt{T}. Figure~\ref{fig:bitweaving} shows
the speedup of \ambit over the baseline for various values of \emph{b}
and \emph{r}.

We draw three conclusions. First, \ambit improves the performance of
the query by between 1.8X and 11.8X (7.0X on average) compared to the
baseline for various values of \emph{b} and \emph{r}. Second, the
performance improvement of \ambit increases with increasing number of
bits per column (\emph{b}), because, as \emph{b} increases, the
fraction of time spent in performing the bitcount operation
reduces. As a result, a larger fraction of the execution time can be
accelerated using \ambit. Third, for \emph{b} = 4, 8, 12, and 16, we
observe large jumps in the speedup of \ambit as we increase the row
count. These large jumps occur at points where the working set stops
fitting in the on-chip cache. By exploiting the high bank-level
parallelism in DRAM, \ambit can outperform the baseline (by up to
4.1X) \emph{even when} the working set fits in the cache.

\subsection{Bitvectors vs. Red-Black Trees}
\label{sec:bitset}

Many algorithms heavily use the \emph{set} data structure. Red-black
trees~\cite{red-black-tree} (RB-trees) are typically used to implement
a set~\cite{stl}. However, a set with a limited domain can be
implemented using a bitvector---a set that contains only elements from
$1$ to $N$, can be represented using an $N$-bit bitvector (e.g.,
Bitset~\cite{stl}). Each bit indicates whether the corresponding
element is present in the set.  Bitvectors provide constant-time
\emph{insert} and \emph{lookup} operations compared to $O(\log n)$
time taken by RB-trees. However, set operations like \emph{union},
\emph{intersection}, and \emph{difference} have to scan the
\emph{entire} bitvector regardless of the number of elements actually
present in the set. As a result, for these three operations, depending
on the number of elements in the set, bitvectors may outperform or
perform worse than RB-trees. With support for fast bulk bitwise
operations, we show that \ambit significantly shifts the trade-off in
favor of bitvectors for these three operations.

\begin{figure}[h]
  \centering
  \includegraphics[scale=1.5]{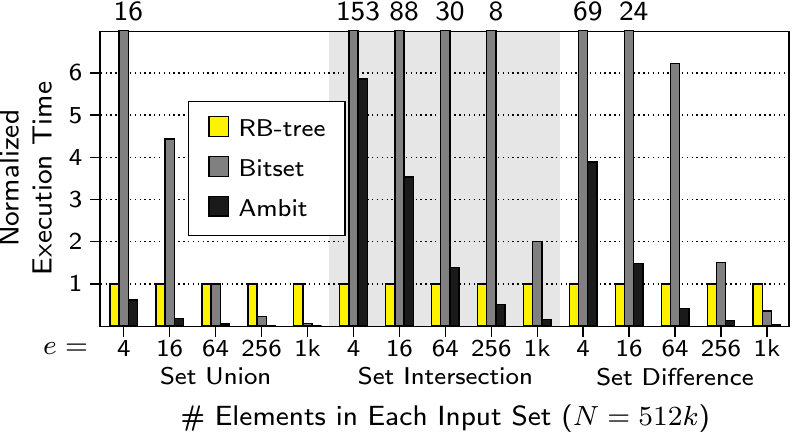}
  \caption{Performance of set operations}
  \label{plot:set-results}
\end{figure}

To demonstrate this, we compare the performance of set \emph{union},
\emph{intersection}, and \emph{difference} using: RB-tree, bitvectors
with 128-bit SIMD support (Bitset), and bitvectors with \ambit. We run
a benchmark that performs each operation on $m$ input sets and stores
the result in an output set. We restrict the domain of the elements to
be from $1$ to $N$. Therefore, each set can be represented using an
$N$-bit bitvector. For each of the three operations, we run multiple
experiments varying the number of elements (\emph{e}) \emph{actually}
present in each input set. Figure~\ref{plot:set-results} shows the
execution time of RB-tree, Bitset, and \ambit normalized to RB-tree
for the three operations for $m = 15$, and $N = 512k$.

We draw three conclusions. First, by enabling much higher throughput
for bulk bitwise operations, \ambit outperforms the baseline Bitset on
all the experiments. Second, as expected, when the number of elements
in each set is very small (16 out of $512k$), RB-Tree performs better
than Bitset and \ambit (with the exception of \emph{union}). Third,
even when each set contains only 64 or more elements out of $512k$,
\ambit significantly outperforms RB-Tree, 3X on average. We conclude
that \ambit makes the bitvector-based implementation of a set more
attractive than the commonly-used red-black-tree-based implementation.

\subsection{Other Applications}
\label{sec:other-apps}

We describe five other examples of applications that can
significantly benefit from \ambit in terms of both performance and
energy efficiency. We leave the evaluation of these applications
with \ambit to future works.

\subsubsection{BitFunnel:  Web Search}

Web search is an important workload in modern systems. From the
time a query is issued to the time the results are sent back, web
search involves several steps. Document filtering is one of the
most time consuming steps. It identifies all documents that
contain all the words in the input query. Microsoft recently
open-sourced BitFunnel~\cite{bitfunnel}, a technology that
improves the efficiency of document filtering. BitFunnel
represents both documents and queries as a bag of words using
Bloom filters~\cite{bloomfilter}. It then uses bitwise AND
operations on specific locations of the Bloom filters to
efficiently identify documents that contain all the query
words. With \ambit, this operation can be significantly
accelerated by simultaneously performing the filtering for
thousands of documents.

\subsubsection{Masked Initialization}

\emph{Masked initializations}~\cite{intel-mmx} are very useful in
applications like graphics (e.g., for clearing a specific color in
an image). Masked operations can be represented using bitwise AND
or OR operations. For example, the $i^{th}$ bit of an integer $x$
can be set using $x = x | (1 << i)$. Similarly, the $i^{th}$ bit
can be reset using $x = x \& (1 << i)$. These masks can be
preloaded into DRAM rows and can be used to perform masked
operations on a large amount of data. This operation can be easily
accelerated using \ambit.

\subsubsection{Encryption}

Many encryption algorithms heavily
use bitwise operations (e.g., XOR)~\cite{xor1,xor2,enc1}. \ambit's
support for fast bulk bitwise operations can i)~boost the performance
of existing encryption algorithms, and ii)~enable new encryption
algorithms with high throughput and efficiency.

\subsubsection{DNA Sequence Mapping}

In DNA sequence mapping, prior
works~\cite{dna-algo1,dna-algo2,dna-algo3,dna-algo4,shd,dna-our-algo,bitwise-alignment,gatekeeper,grim,lee2015,nanopore-sequencing,shouji,magnet}
propose algorithms to map sequenced \emph{reads} to the reference
genome. Some
works~\cite{shd,dna-our-algo,bitwise-alignment,gatekeeper,grim,myers1999,shouji}
heavily use bulk bitwise operations. \ambit can significantly improve
the performance of such DNA sequence mapping algorithms~\cite{grim}.

\subsubsection{Machine Learning}

Recent works have shown that deep neural networks (DNNs)
outperform other machine learning techniques in many problems such
as image classification and speech recognition. Given the high
computational requirements of DNNs, there has been a focus on
neural networks with binary values~\cite{xnor-net,bnn}. A recent
work also exploits bit-serial computation to execute DNN inference
algorithms using SIMD bitwise operations in the on-chip SRAM
cache~\cite{neural-cache}. In conjunction with such techniques
that increase the fraction of bitwise operations in DNNs, \ambit
can significantly improve the performance of DNN algorithms.

\section{Future Work}

There are several avenues for future work that can build on top of
the \ambit proposal. We briefly describe these avenues in this
section.

\subsection{Extending \ambit to Other Operations}

We envision two major class of operations that can significantly
increase the domain of applications that can benefit from
\ambit.

The first operation is \emph{count}. Counting the number of non-zero
bits can be a very useful operation in many applications, including
the ones evaluated in this paper
(Section~\ref{sec:applications}). Extending this operation to count
the number of non-zero integers (of different widths) can further
expand the scope of this operation.

The second operation is \emph{shift}. Most arithmetic operations
require some kind of bitwise shift. Shifting is also heavily used
in any encryption algorithms~\cite{xor1,xor2,enc1} and
bioinformatics
algorithms~\cite{bitwise-alignment,shd,gatekeeper,grim,myers1999,shouji}. Extending
\ambit to support bit shifting at different granularities can
allow \ambit to significantly accelerate these applications. While
recent works~\cite{drisa,dracc} explore a possible implementation
of shifting inside DRAM, several challenges remain unaddressed
(e.g., data mapping, handling column failures in DRAM).

\subsection{Evaluation of New Applications with \ambit}

In Section~\ref{sec:other-apps}, we discussed five other applications
that can benefit significantly from \ambit. A concrete evaluation of
these applications can 1)~further strengthen the case for \ambit, and
2)~reveal other potential operations that may require acceleration in
DRAM in order to obtain good end-to-end application speedups (e.g.,
\emph{bitcount}).

\subsection{Redesigning Applications to Exploit \ambit}

In Section~\ref{sec:bitweaving}, we show how the
BitWeaving~\cite{bitweaving} technique can be accelerated using
\ambit. BitWeaving is a technique that \emph{redesigns} the data
layout of tables in a database and uses an appropriate algorithm
to execute large scan queries efficiently. It is the modified data
layout that makes BitWeaving amenable for SIMD/\ambit
acceleration. We believe similar careful redesign techniques can
be used for other important workloads such as graph processing,
machine learning, bioinformatics algorithms, etc., that can make
them very amenable for acceleration using \ambit.

\subsection{Taking Advantage of Approximate \ambit}

\ambit requires a costly ECC mechanism to avoid errors during
computation. In addition, the error rate may increase with
increasing process variation. However, if an application can
tolerate errors in computation, as observed by prior
works~\cite{approx-computing,yixin-dsn,rfvp-taco,rfvp-pact}, then
\ambit can be used as an approximate in-DRAM computation substrate.

\section{Conclusion}

In this paper, we focused on bulk bitwise operations, a class of
operations heavily used by some important applications. Existing
systems are inefficient at performing such operations as they have
to transfer a large amount of data on the memory
channel, resulting in high latency, high memory bandwidth
consumption, and high energy consumption.

We described \ambit, which employs the notion of \emph{Processing
  using Memory} introduced in a recent
work~\cite{pum-bookchapter}. Ambit converts DRAM-based main memory
into a bulk bitwise operation execution engine that performs bulk
bitwise operations completely inside the
memory~\cite{ambit}. Ambit has two component mechanisms. The first
mechanism exploits the fact that many DRAM cells share the same
sense amplifier and uses the idea of simultaneous activation of
three rows of DRAM cells to perform bitwise MAJORITY/AND/OR operations
efficiently. The second mechanism uses the inverters already
present inside DRAM sense amplifiers to efficiently perform
bitwise NOT operations. Since Ambit heavily exploits the internal
organization and operation of DRAM, it incurs very low cost on top
of the commodity DRAM architecture ($< 1\%$ chip area overhead).

Our evaluations show that Ambit enables between one to two order
of magnitude improvement in raw throughput and energy consumption
of bulk bitwise operations compared to existing DDR interfaces. We
describe many real-world applications that can take advantage of
Ambit. Our evaluations with three such applications show that
Ambit can improve average performance of these applications
compared to the baseline by between 3.0X-11.8X. Given its low cost and
large performance improvements, we believe Ambit is a promising
execution substrate that can accelerate applications that heavily
use bulk bitwise operations. We hope future work will uncover even
more potential in Ambit like execution.

\section*{Acknowledgments}

We thank the reviewers of ISCA 2016/2017, MICRO 2016/2017, and
HPCA 2017 for their valuable comments on various drafts of shorter
versions of this work. We thank the members of the SAFARI group
and PDL for their feedback. We acknowledge the generous support of
our industrial partners, over the years, especially AliBaba,
Google, Huawei, Intel, Microsoft, Nvidia, Samsung, Seagate, and
VMWare. This work was supported in part by NSF, SRC, and the Intel
Science and Technology Center for Cloud Computing. Some components
of this work appeared in IEEE CAL~\cite{bitwise-cal},
MICRO~\cite{ambit}, ADCOM~\cite{pum-bookchapter}, and Seshadri's
Ph.D. thesis~\cite{vivek-thesis}.

\bibliographystyle{plain}
\bibliography{references}

\end{document}